\documentclass[prl,twocolumn,letterpaper,superscriptaddress]{revtex4-1}
\usepackage[space]{grffile}
\usepackage{bm,graphicx,graphics,amsmath,amssymb,bm,epsfig,color}
\usepackage{euscript,multirow,tabularx}
\usepackage{longtable}
\usepackage{pifont}
\usepackage{xcolor}
\usepackage{float}  
\usepackage{changes}
\usepackage{ulem}
\begin{document}


\title{Short-range Thermal Magnon Diffusion in Magnetic Garnet} 

\author{K. An}
\affiliation{Université Grenoble Alpes, CEA, CNRS, Grenoble INP, Spintec, 38054 Grenoble, France}

\author{R. Kohno}
\affiliation{Université Grenoble Alpes, CEA, CNRS, Grenoble INP, Spintec, 38054 Grenoble, France}

\author{N. Thiery}
\affiliation{Université Grenoble Alpes, CEA, CNRS, Grenoble INP, Spintec, 38054 Grenoble, France}

\author{D. Reitz}
\affiliation{Department of Physics and Astronomy, University of California, Los Angeles, California 90095, USA}

\author{L. Vila}
\affiliation{Université Grenoble Alpes, CEA, CNRS, Grenoble INP, Spintec, 38054 Grenoble, France}

\author{V. V. Naletov} 
\affiliation{Université Grenoble Alpes, CEA, CNRS, Grenoble INP, Spintec, 38054 Grenoble, France}
\affiliation{Institute of Physics, Kazan Federal University, Kazan
    420008, Russian Federation}

\author{N. Beaulieu} 
\affiliation{SPEC, CEA-Saclay, CNRS, Universit\'e Paris-Saclay,
  91191 Gif-sur-Yvette, France}
\affiliation{LabSTICC, CNRS, Universit\'e de Bretagne Occidentale,
  29238 Brest, France}

\author{J. Ben Youssef} 
\affiliation{LabSTICC, CNRS, Universit\'e de Bretagne Occidentale,
  29238 Brest, France}

\author{G. de Loubens} 
\affiliation{SPEC, CEA-Saclay, CNRS, Universit\'e Paris-Saclay,
  91191 Gif-sur-Yvette, France}

\author{Y. Tserkovnyak}
\affiliation{Department of Physics and Astronomy, University of California, Los Angeles, California 90095, USA}

\author{O. Klein}
\email[Corresponding author:]{ oklein@cea.fr}
\affiliation{Université Grenoble Alpes, CEA, CNRS, Grenoble INP, Spintec, 38054 Grenoble, France}

\date{\today}

\begin{abstract}
Using the spin Seebeck effect (SSE), we study the propagation distance of thermal spin currents inside a magnetic insulator thin film in the short-range regime. We disambiguate spin currents driven by temperature and chemical potential gradients by comparing the SSE signal before and after adding a thermalization capping layer on the same device. We report that the measured spin decay behavior near the heat source is well accounted for by a diffusion model where the magnon diffusion length is in submicron range, \textit{i.e.} two orders of magnitude smaller than previous estimates inferred from the long-range behavior. Our results highlight the caveat in applying a diffusive theory to describe thermal magnon transport, where a single decay length may not capture the behavior on all length scales.
\end{abstract}

\maketitle

The generation of pure spin currents by heat \cite{bauer2012spin,boona2014spin} is a tantalizing issue, which offers a unique opportunity to reach strong out-of-equilibrium regime with large spin current density produced inside a magnetic material \cite{weiler2013experimental}. Interests lie in the prospect of reaching new collective dynamical behaviors of spin transport such as the hydrodynamic regime conspicuous by the emergence of turbulences \cite{ulloa2019nonlocal}. Magnon superfluidity may even establish when the density exceeds the Bose-Einstein condensation threshold under large temperature gradients applied to low damping magnetic insulators, such as yttrium iron garnets (YIG) \cite{bender2012electronic,tserkovnyak2016bose}, where local heating can be provided by injecting a large electrical current density through an adjacent metal, advantageously in Pt \cite{safranski2017spin,thiery2018nonlinear,schneider2020bose}, or by optically heating with a laser \cite{weiler2012local,jamison2019long,olsson2020pure}.

The spin transport properties are governed by $\lambda$, the characteristic decay length of the spin information. Previous reports on measuring $\lambda$ in YIG at room temperature by the spin Seebeck effect (SSE) indicates that for distances larger than $\sim$10\,$\mu$m (long-range regime), the SSE signal follows an exponential decay with a characteristic length of the order of $\lambda_0 \approx 10~\mu$m  \cite{cornelissen2015long,giles2017thermally}. This large value is believed to exceed $\ell$, the magnon mean free path (energy non-conserving decay length), which is expected to be on the order of few nanometers \cite{boona2014magnon,Dyson1956}, suggesting there that magnons behave as a diffusive fluid \cite{flebus2016two}. However such a large value of $\lambda_0$ may seem surprising for thermal magnons. If one extends the magnon dispersion up to the THz-range, the extrapolated ballistic decay length for thermal magnons is $\lambda_\text{bal}=\lambda_\text{ex} /(2 \alpha \sqrt{\omega_T / \omega_M})=2$~$\mu$m, where $\omega_T = k_B T_0/ \hbar = 2 \pi \times 6.25$~THz, $T_0$ is room temperature, $\lambda_\text{ex} \approx 15$~nm is the exchange length in YIG, $\alpha \approx 10^{-4}$ \footnote{$\alpha \sim 10^{-4}$ is a lower bound for the Gilbert damping in YIG thin films with thickness $<$~100nm} is the Gilbert damping, and $\omega_M = \gamma \mu_0 M_s = 2 \pi \times 4.48$~GHz. However this estimate $\lambda_\text{bal}$, which is already smaller than $\lambda_0$, should be considered as an upper bound because \textit{i)} the magnon lifetime is expected to be reduced in the THz-range \cite{cherepanov1993saga} \textit{ii)} the group velocity is reduced towards the edge of the Brillouin zone \cite{plant1977spinwave,barker2016thermal}, and \textit{iii)} it does not account for the  $\sqrt{\ell/\lambda_\text{bal}}$ reduction of the characteristic propagation distance due to diffusion process. 

In fact, the distance range of the transport study is also a potent mean to select a very specific part of the magnon\deleted{s} spectrum. In experiments focusing on the long-range behavior, one has in essence efficiently filtered out any short decay magnons. Behind this debate lies a fundamental question of how well magnon transport can be described by a diffusive model forming one fluid with a single $\lambda$, whose value would govern SSE on all length scales. Submicron lengths have been inferred from several longitudinal SSE measurements in spatial \cite{kehlberger2015length,prakash2018evidence} and time domain \cite{agrawal2014role,jamison2019long}. In nonlocal SSE measurements, where two different Pt strips are used for the spin injection and detection, only longer spin decay lengths have been reported. The existence of shorter decay lengths has been difficult to observe because the voltage induced by SSE shows a nontrivial spatial decay as a function of the Pt detector position near the heat source \cite{shan2016influence,ulloa2019nonlocal}. The complex decay profile can be attributed to the competition between magnons driven by the gradients of temperature and magnon chemical potential \cite{adachi2013theory,flebus2016two,cornelissen2016magnon}. It has been difficult to control these two sources of spin excitation in experiments, which hinders a correct extraction of a characteristic decay length near the heat source.

\begin{figure}
    \includegraphics[width=0.5\textwidth]{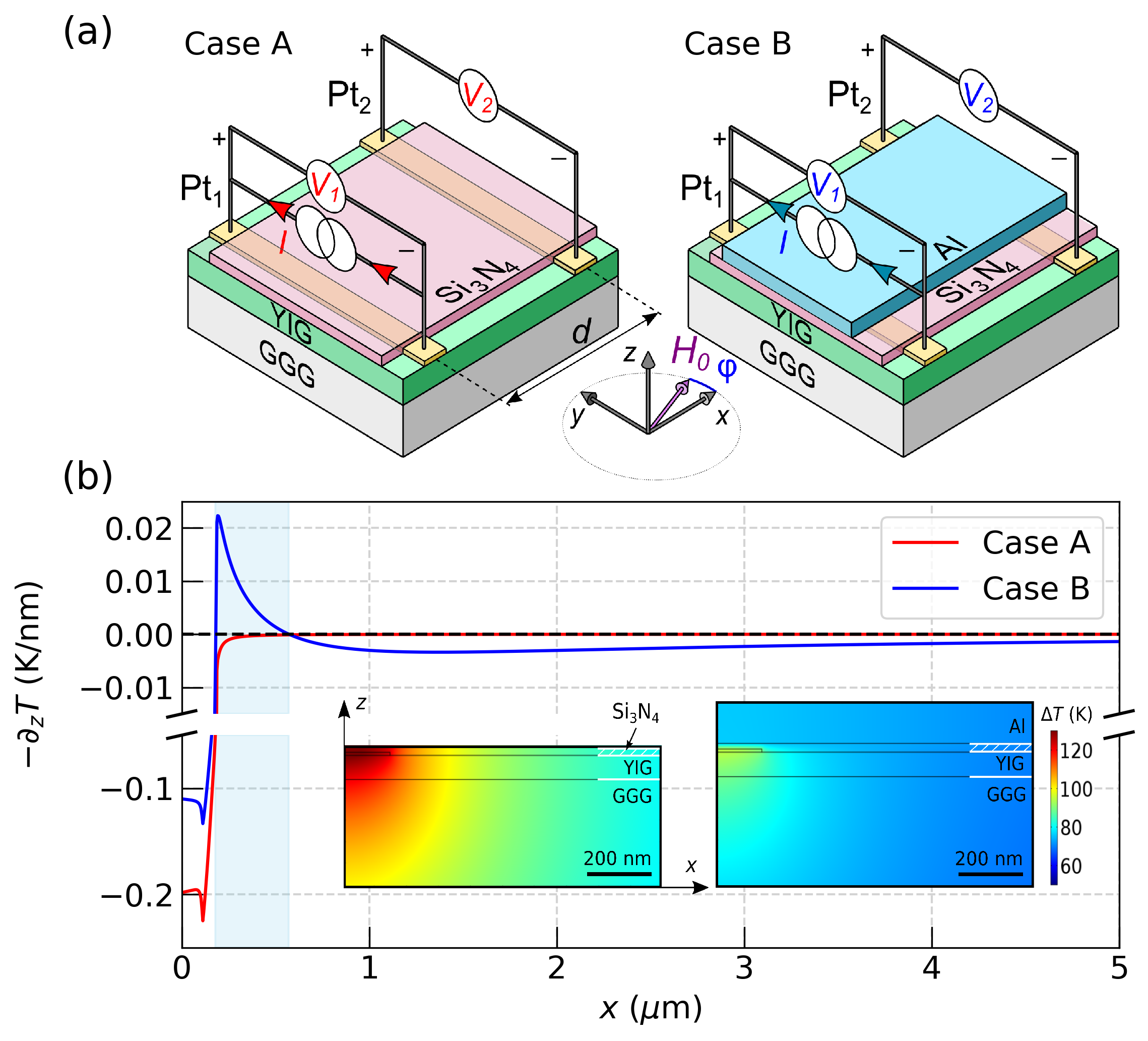}
    \caption{(a) Comparison of the measurements of the local ($V_1$) and nonlocal ($V_2$) voltages generated in YIG$|$Pt$|$Si$_3$N$_4$ (case A) and YIG$|$Pt$|$Si$_3$N$_4|$Al (case B) stacks. Experiments are performed on the same devices before and after the deposition of an Al capping layer (left and right schematics). Two Pt electrodes deposited on top of YIG film  monitor the spin-transconductance when an external magnetic field $H_0$ rotates in-plane in the azimuthal direction $\varphi$. The center-to-center distance $d$ between the two electrodes is varied between 0.5\,$\mu$m to 6.3\,$\mu$m. (b) Calculated vertical temperature gradient profiles at the top YIG surface at 2 mA. The light blue shaded region indicates the inverted gradient in case B. Beyond this region, $\partial_z T$ is about three orders of magnitude larger in case B. The insets are the calculated temperature profiles for both cases.}
    \label{Fig1}
\end{figure}

In this paper we develop a way to disambiguate these two contributions after altering the temperature profile. We monitor on the same devices the short-range SSE signal before (case A) and after (case B) capping it with a non-magnetic aluminum layer. The capping allows to change the vertical thermal gradient without altering the YIG interface. We observe that the sign of SSE voltage inverts twice within a distance of $1$~$\mu$m from the heat source for case B. The corresponding sign reversal of SSE suggest that the magnons clearly sense the change in local temperature gradient taking place for case B. With a simple diffusive transport model, the measured SSE decay profile for both cases can be reproduced if one introduces a thermal magnon diffusion length $\lambda \approx 300 \pm 200$~nm. The extracted short $\lambda$ from our measurement fills the gap between different length scales reported in the longitudinal and nonlocal SSE measurements.


We use a 56\,nm thick YIG(111) film grown on a 500$\,\mu$m GGG substrate by liquid phase epitaxy. Ferromagnetic resonance experiments have shown a damping parameter of  $4\, \times$10$^{-4}$ revealing an excellent crystal quality of the YIG film \cite{beaulieu2018temperature}. The sample structure and measurement configuration are shown in Fig. \ref{Fig1}(a). In our notation, subscripts 1 and 2 refer to the voltages measured by the Pt$_1$ and Pt$_2$, respectively. We show the data for both YIG$|$Pt$|$Si$_3$N$_4$ (case A, red) and YIG$|$Pt$|$Si$_3$N$_4|$Al (case B, blue). The color conventions will be used consistently throughout the paper. Two Pt strips (Pt$_1$ and Pt$_2$) with width  of 300$\,$nm, length of 30$\,\mu$m and thickness of 7\,nm have been evaporated directly on top of the YIG film. The center-to-center distance $d$ between two Pt strips varies from 0.5 to 6.3\,$\mu$m. The sample is then covered by a 20\,nm thick Si$_3$N$_4$ protection film. After full characterization of the different devices, a 105$\,$nm thick Aluminum layer with length 30\,$\mu$m and width 10 $\mu$m is deposited on the top of the Si$_3$N$_4$ film, and the same devices are measured again. The sample is submitted to an external field of $\mu_0 H_0\,=\,200\,$mT rotating within the $xy$ plane (in-plane configuration).


We first show the expected change in temperature profile by the Al capping in Fig. \ref{Fig1}(b). At the Pt injector, $\Delta T$ is about 40 K lower in case B. The reduced temperature rise is experimentally confirmed by measuring the Pt$_1$ resistance (see the supplemental material \cite{suppl}). Besides the change in the temperature profile, the gradient profile also shows a dramatic change. While in case A the thermal gradient is always directed downwards (into the substrate), in case B, a large thermal gradient directed upwards (into Al) is created half a micron away from the source. The shaded region highlights the effect. Since the vertical thermal gradient drives the SSE, this feature gives rise to an additional signal at Pt$_2$.


\begin{figure}
    \includegraphics[width=0.5\textwidth]{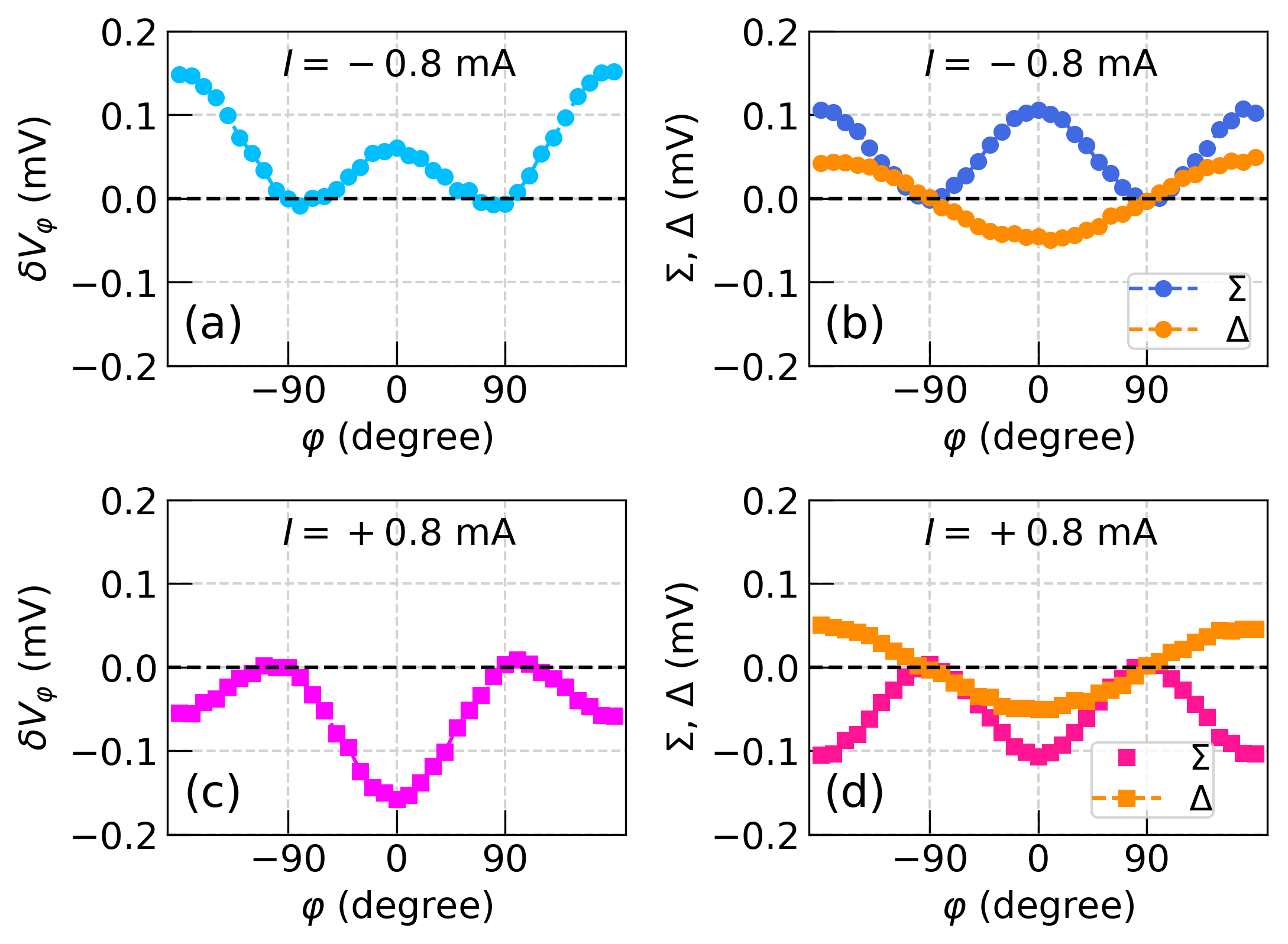}
    \caption{(a,c) Angular dependence of the background subtracted local voltage $\delta V_\varphi$ measured in the Pt injector strip for the current of $I=\pm\, 0.8\;$mA and external magnetic field of $\mu_0 H_0=200\;$mT. We subtract the reference voltage $V_{y}$ from the raw signal to remove any contributions not associated with magnons. In (b) and (d) we decompose the measured magnetoresistive voltage into two components : $\Sigma$ and $\Delta$, the even and odd contributions of the signal with respect to the mirror symmetry along the $yz$ plane (see the main text).} 
    \label{Fig2}
\end{figure}
 
\begin{figure*}
    \includegraphics[width=1\textwidth]{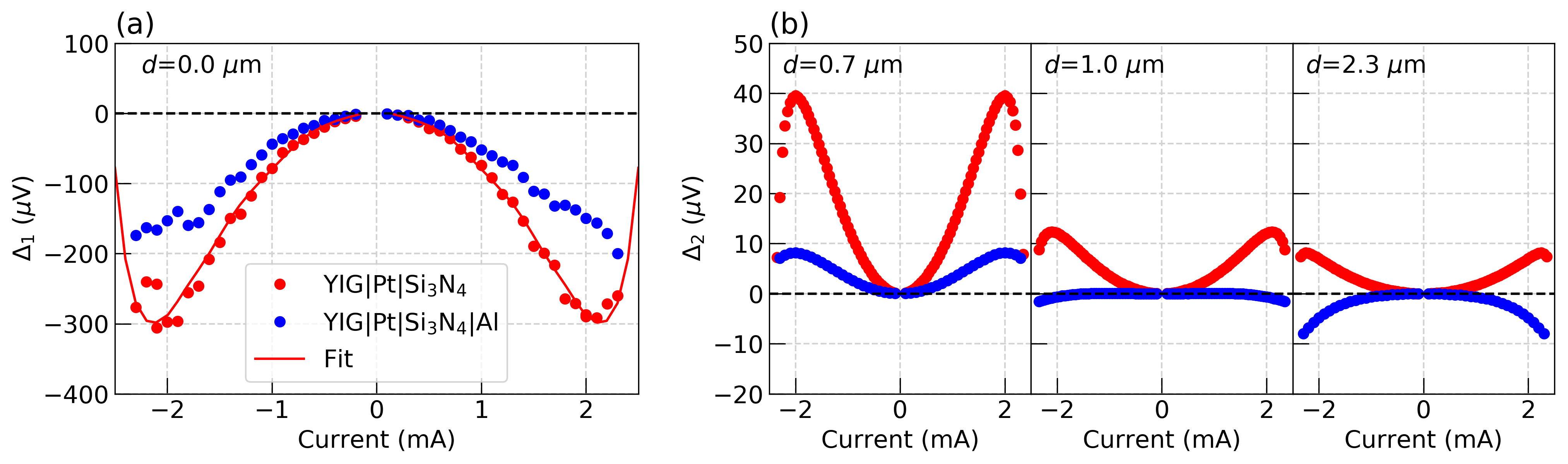}
    \caption{(a) Current dependence of measured local $\Delta_1$ for case A (YIG$|$Pt$|$Si$_3$N$_4$) and case B (YIG$|$Pt$|$Si$_3$N$_4|$Al). The red solid line shows that the local $\Delta_1$ follows the expected behavior based on the temperature rise with increasing current. (b) Measured current dependence of nonlocal $\Delta_2$ for three different $d$'s for the case A (red) and B (blue).}
    \label{Fig3}
\end{figure*}

We use the same method demonstrated in our previous work to extract the SSE voltage \cite{thiery2018nonlinear}. As an illustration, we display the background-subtracted local voltage $\delta V_\varphi = V_\varphi - V_y$ measured with the Pt injector on YIG$|$Pt$|$Si$_3$N$_4$ at $\pm$\,0.8$\,$mA as a function of the in-plane magnetic field angle $\varphi$ in Fig. \ref{Fig2}(a). The offset $V_{y}$, measured when the magnetic field is applied along the $y$ axis ($\varphi$ = 90$^{\circ}$), takes account of all the spurious contributions to the spin transport \cite{thiery2018electrical}. To distinguish the SSE from the spin orbit torque, we define two quantities based on the {$yz$}{} mirror symmetry: $\Sigma_{\varphi,I},\Delta_{\varphi,I}  \equiv \left ( \delta V_{\varphi,I} \pm \delta V_{\overline \varphi ,I} \right ) /{2}$
where $\overline \varphi = \pi-\varphi $. Figure \ref{Fig2}(b) and (c) show the evolution of the extracted $\Sigma$ and $\Delta$ as a function of $\varphi$ for both polarities of the current $I$. $\Sigma$ is antisymmetric with respect to the current and evolves as $\cos 2\varphi$, as expected from the spin Hall magnetoresistance effect \cite{chen2013theory,nakayama2013spin,hahn2013comparative}. $\Delta$ shows a $\cos \varphi$ angular dependence and is symmetric with respect to the current, consistent with the SSE \cite{uchida2008observation}. In the following, we shall exclusively focus on the SSE voltage ($\Delta$). 


Next, we compare the full current dependence of the SSE voltage $\Delta_1(I)$ ($\Delta_{\varphi=0,I}$ in Pt$_1$) for both case A (red) and case B (blue) in Fig. \ref{Fig3}(a). We clearly see that the voltage is negative for both cases over the entire current range. The parabolic curvature observed at low currents decreases when the Al heat sink is introduced, which agrees with the reduced temperature rise. We observe that $\Delta_1(I)$  reaches a minimum at 2~mA for case A with the minimum shifting to a higher current for case B. We attribute this reversal of the slope as the growing influence of the vanishing YIG magnetization as one approaches $T_c$=560 K, the Curie temperature of YIG \cite{gilleo1958magnetic}. The current dependence of $\Delta_1(I)$ for case A can be reproduced by an empirical formula $\Delta_1 \propto M(T) (T-T_0)$, where $M(T)$ is the saturation magnetization at the temperature $T$, while $T_0=300$~K is the temperature of the substrate. From the fit as shown in Fig. \ref{Fig3}(a) one can extract the spin Seebeck coefficient 0.08\,$\mu$V\,K$^{-1}$ in good agreement with previous estimates \cite{guo2016influence} (see the supplemental material for detail \cite{suppl}).

\begin{figure*}
    \includegraphics[width=1\textwidth]{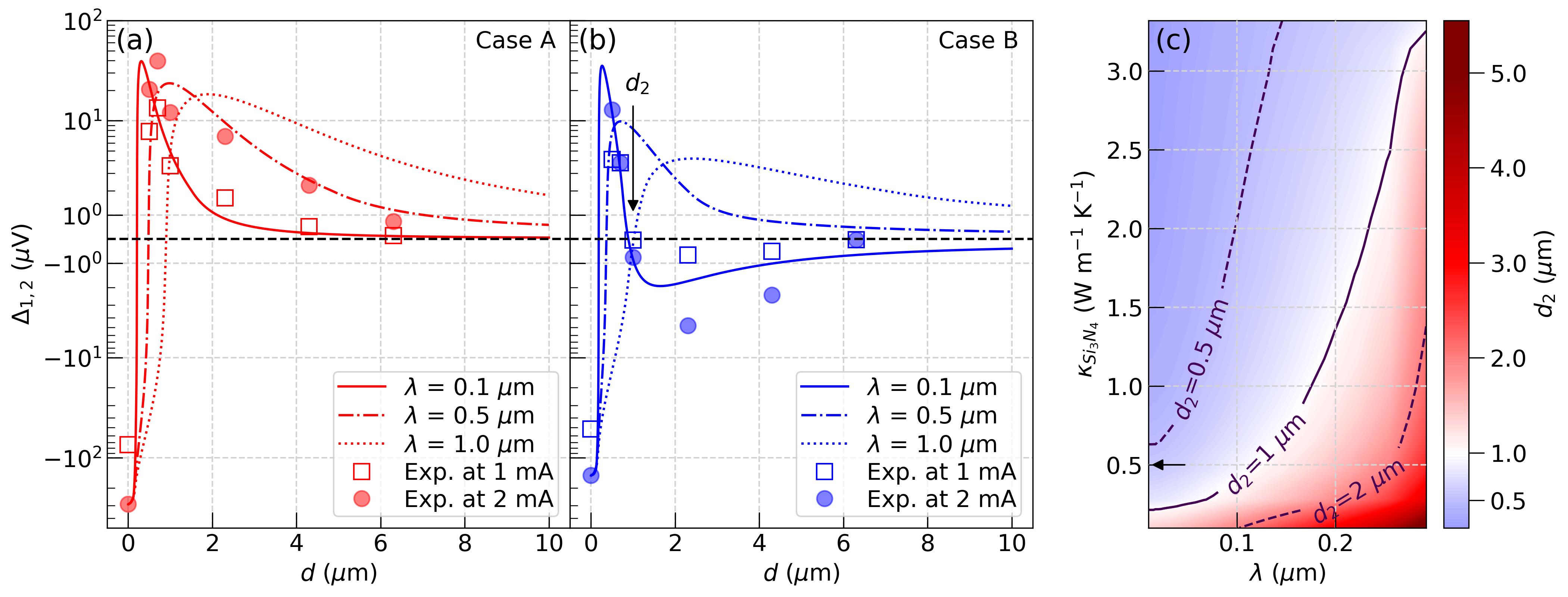}
    \caption{Comparison of the experimentally measured $\Delta$'s at 1 and 2 mA with the numerically calculated chemical potential profile for (a) case A and (b) case B on a symmetric log scale. The result with $\lambda$ = 0.1 $\mu$m reproduces the measured double crossing as shown in (b). The second crossing does not appear anymore when increasing $\lambda$ to 0.5 or 1\,$\mu$m. (c) Contour plot of the second crossing position $d_2$ as a function of $\lambda$ and thermal conductivity of Si$_3$N$_4$. The black lines represent the iso-lines for different $d_2$'s. The arrow points to the value of $\kappa_\text{Si$_3$N$_4$}$ used in panel (a) and (b).}
    \label{Fig4}
\end{figure*}

While the local voltage $\Delta_1$ is negative, the non-local voltage $\Delta_2$'s from devices in case A are positive for all the different distances (red curves in Fig. \ref{Fig3}(b)). This observation is consistent with the previous works that reported a single sign reversal of the SSE voltage measured as a function of distance from the heat source \cite{shan2016influence,ganzhorn2017temperature,zhou2017lateral}. It has been reported that the characteristic distance at which the magnon accumulation changes the sign can be tuned by varying the magnetic film thickness \cite{shan2016influence} or magnon diffusion length \cite{ganzhorn2017temperature}. 
The situation is quite different for case B, as shown with the blue dots in Fig. \ref{Fig3}(b). In the vicinity of the injector, the nonlocal $\Delta_2$ is still positive at 0.7\,$\mu$m but much smaller than case A. The sign of $\Delta_2$ for case B eventually becomes negative when the Pt detectors are positioned at 1\,$\mu$m and 2.3\,$\mu$m away from the injector. Our understanding of the sign changes as a function of $d$ is as follows: at $d = 0$, heat drags magnons from Pt$_1$ down into YIG (negative sign SSE); at larger $d$, magnons can transfer spin from the YIG bulk into Pt$_2$ (positive sign); adding Al creates a region of inverted heat flow over a short range (positive sign within 0.2 and 0.5 $\mu$m) and significant heat flow down into YIG over a longer range (negative sign). The fact that we see a negative signal after the second crossing signifies that the positive SSE driven by the magnon diffusion already becomes diminished and the local temperature driven SSE dominates. In this perspective the second crossing point may put a constraint on the estimate of the thermal magnon diffusion length.

To model our experiment, we treat the magnons in YIG as a diffusive gas with temperature $T$ and chemical potential $\mu$ described by a set of transport coefficients \cite{flebus2016two}. The measured $\Delta$ signal is proportional to the magnon chemical potential at the interface between YIG and Pt \cite{castel2012platinum}. The continuity equation for spin current density $J_s$ in the steady state is 
\begin{equation}
    -\boldsymbol{\nabla} \cdot \boldsymbol{J_s}= g \mu,
    \label{eq1}
\end{equation}
where $g$ is spin relaxation coefficient. In linear response, the transport equation is
\begin{equation}
    \boldsymbol{J_s}=-\sigma (\boldsymbol{\nabla} \mu + \varsigma \boldsymbol{\nabla} T),
    \label{eq2}
\end{equation}
where $\sigma$ is the spin conductivity and $\varsigma$ is the bulk spin Seebeck coefficient. The two equations are combined and lead to
\begin{equation}
    \frac{\mu}{\lambda^2} = \nabla^2\mu + \varsigma \nabla^2 T ,
    \label{eq3}
\end{equation}
where $ \lambda=\sqrt{\sigma/g}$ is the thermal magnon diffusion length. 

We use a finite element method, COMSOL, to calculate the temperature profile and the magnon chemical potential $\mu$ in a 2D geometry (see supplementary material for details \cite{suppl}). The second crossing point $d_2$ observed in the case B, as indicated in Fig. \ref{Fig4}(b), is an important feature that reveals the inverted heat flow near the heat source as shown in Fig. \ref{Fig1}(b). The calculated spatial profiles of $\mu$ for three different values of $\lambda$ are compared with the experimental data in Figs. \ref{Fig4}(a) and (b) after normalization to the measured values at $d$ = 0 $\mu$m. We find that the second sign reversal at $d_2$ can be reproduced with a short $\lambda=100$~nm. Increasing it to 0.5 or 1 $\mu$m, however, no longer reproduces the second crossing in case B (compare solid line with dashed and dotted lines in Fig. \ref{Fig4}(b)).

The fit value of $\lambda$ depends on other parameters in the model. The second crossing point, $d_2$, can be affected by the temperature profile at the interface, which is sensitive for example to the thermal conductivity of Si$_3$N$_4$. We show a contour plot for $d_2$ as a function of $\kappa_\text{Si$_3$N$_4$}$ and $\lambda$ in Fig. \ref{Fig4}(c). The measured second crossing constraints the parameter space to the line along $d_2$ = 1 $\mu$m. The second crossing is not observed for $\lambda >$ 500 nm regardless of $\kappa_\text{Si$_3$N$_4$}$. These considerations can be further modified by the interfacial SSE due to the magnon-phonon temperature difference at the interface \cite{xiao2010theory,bender2015interfacial}. Enhanced local temperature driven contribution to the measured signal can increase $\lambda$ for a given $d_2$. {To have \emph{quantitative} agreement with the data, we get there an upper bound of also $\lambda \sim$\,500$~$nm, although it is worth to mention that in the limit of large temperature mismatch between the YIG and the Pt one can reproduce but only \emph{qualitatively} the single crossing in case A and double crossing in case B for much larger $\lambda$ (see the supplemental material \cite{suppl}).}


Although our proposed fit with a diffusive equation parametrized by $\lambda \in [100,500] \approx 300\pm200$~nm captures well the short-range behavior in both cases A and B, we emphasize that this does not contradict earlier works \cite{cornelissen2015long,shan2016influence}. The data observed for case A are similar to the one already observed in other YIG devices, where the fit of the long-range decay behavior has lead to the larger $\lambda_0 \approx 30 \times \lambda$ \footnote{Our pulse method, which allows the injection of large current densities in the Pt, does not have the dynamical range to follow the SSE signal at distances above 10~$\mu$m, where the signal is diminished by three orders of magnitude. This prevents us from studying if there is another tail at this exponential decay with a characteristic length scale of about tens of micron as reported in other studies.}. Therefore the difference is not the illustration of a discrepancy but rather of the limit of the diffusive model. We also note that our extracted $\lambda$ is close to the previously reported energy relaxation length of magnons \cite{agrawal2014role,kehlberger2015length,prakash2018evidence,jamison2019long}, where the connection needs to be further explored \footnote{Probably the energy relaxation length of magnons is also a multiscale issue}. Also being linear our theory does not explain why the decay length seems to depend on the current values (filled dots vs unfilled squares in Fig. \ref{Fig4}(a)). However it has been established that at large current densities, one is approaching the damping compensation for subthermal magnons \cite{thiery2018nonlinear}, a mechanism that would significantly increase their populations. Significant participation of these magnons to the SSE should increase the decay length, in agreement with the experimental findings observed at higher current \cite{thiery2018nonlinear}. This suggests that low energy magnons are also involved in the SSE \cite{jamison2019long}.

As a summary, we measure the spatial distribution of thermal magnons in a thin YIG film. We altered the temperature profile across the YIG film with an aluminum layer. This results in that the non-equilibrium thermal magnon profile deviates from an exponential decay and shows a double sign reversal. We use a linear response magnon transport theory to obtain the short-range thermal magnon diffusion length of a submicron range, which is about two orders of magnitude smaller than the value found in previous reports focused on the long-range measurements. Our results suggest that local heating produces a multiscale magnon diffusion phenomena, where magnons at thermal energy remain mostly localized near the source.

\begin{acknowledgments}

\end{acknowledgments}

\bibliography{bib}

\begin{thebibliography}{43}%
\makeatletter
\providecommand \@ifxundefined [1]{%
 \@ifx{#1\undefined}
}%
\providecommand \@ifnum [1]{%
 \ifnum #1\expandafter \@firstoftwo
 \else \expandafter \@secondoftwo
 \fi
}%
\providecommand \@ifx [1]{%
 \ifx #1\expandafter \@firstoftwo
 \else \expandafter \@secondoftwo
 \fi
}%
\providecommand \natexlab [1]{#1}%
\providecommand \enquote  [1]{``#1''}%
\providecommand \bibnamefont  [1]{#1}%
\providecommand \bibfnamefont [1]{#1}%
\providecommand \citenamefont [1]{#1}%
\providecommand \href@noop [0]{\@secondoftwo}%
\providecommand \href [0]{\begingroup \@sanitize@url \@href}%
\providecommand \@href[1]{\@@startlink{#1}\@@href}%
\providecommand \@@href[1]{\endgroup#1\@@endlink}%
\providecommand \@sanitize@url [0]{\catcode `\\12\catcode `\$12\catcode
  `\&12\catcode `\#12\catcode `\^12\catcode `\_12\catcode `\%12\relax}%
\providecommand \@@startlink[1]{}%
\providecommand \@@endlink[0]{}%
\providecommand \url  [0]{\begingroup\@sanitize@url \@url }%
\providecommand \@url [1]{\endgroup\@href {#1}{\urlprefix }}%
\providecommand \urlprefix  [0]{URL }%
\providecommand \Eprint [0]{\href }%
\providecommand \doibase [0]{http://dx.doi.org/}%
\providecommand \selectlanguage [0]{\@gobble}%
\providecommand \bibinfo  [0]{\@secondoftwo}%
\providecommand \bibfield  [0]{\@secondoftwo}%
\providecommand \translation [1]{[#1]}%
\providecommand \BibitemOpen [0]{}%
\providecommand \bibitemStop [0]{}%
\providecommand \bibitemNoStop [0]{.\EOS\space}%
\providecommand \EOS [0]{\spacefactor3000\relax}%
\providecommand \BibitemShut  [1]{\csname bibitem#1\endcsname}%
\let\auto@bib@innerbib\@empty
\bibitem [{\citenamefont {Bauer}\ \emph {et~al.}(2012)\citenamefont {Bauer},
  \citenamefont {Saitoh},\ and\ \citenamefont {Van~Wees}}]{bauer2012spin}%
  \BibitemOpen
  \bibfield  {author} {\bibinfo {author} {\bibfnamefont {G.~E.}\ \bibnamefont
  {Bauer}}, \bibinfo {author} {\bibfnamefont {E.}~\bibnamefont {Saitoh}}, \
  and\ \bibinfo {author} {\bibfnamefont {B.~J.}\ \bibnamefont {Van~Wees}},\
  }\bibfield  {title} {\bibinfo {title} {Spin caloritronics},\ }\href@noop {}
  {\bibfield  {journal} {\bibinfo  {journal} {Nature materials}\ }\textbf
  {\bibinfo {volume} {11}},\ \bibinfo {pages} {391} (\bibinfo {year}
  {2012})}\BibitemShut {NoStop}%
\bibitem [{\citenamefont {Boona}\ \emph {et~al.}(2014)\citenamefont {Boona},
  \citenamefont {Myers},\ and\ \citenamefont {Heremans}}]{boona2014spin}%
  \BibitemOpen
  \bibfield  {author} {\bibinfo {author} {\bibfnamefont {S.~R.}\ \bibnamefont
  {Boona}}, \bibinfo {author} {\bibfnamefont {R.~C.}\ \bibnamefont {Myers}}, \
  and\ \bibinfo {author} {\bibfnamefont {J.~P.}\ \bibnamefont {Heremans}},\
  }\bibfield  {title} {\bibinfo {title} {Spin caloritronics},\ }\href@noop {}
  {\bibfield  {journal} {\bibinfo  {journal} {Energy \& Environmental Science}\
  }\textbf {\bibinfo {volume} {7}},\ \bibinfo {pages} {885} (\bibinfo {year}
  {2014})}\BibitemShut {NoStop}%
\bibitem [{\citenamefont {Weiler}\ \emph {et~al.}(2013)\citenamefont {Weiler},
  \citenamefont {Althammer}, \citenamefont {Schreier}, \citenamefont {Lotze},
  \citenamefont {Pernpeintner}, \citenamefont {Meyer}, \citenamefont {Huebl},
  \citenamefont {Gross}, \citenamefont {Kamra}, \citenamefont {Xiao} \emph
  {et~al.}}]{weiler2013experimental}%
  \BibitemOpen
  \bibfield  {author} {\bibinfo {author} {\bibfnamefont {M.}~\bibnamefont
  {Weiler}}, \bibinfo {author} {\bibfnamefont {M.}~\bibnamefont {Althammer}},
  \bibinfo {author} {\bibfnamefont {M.}~\bibnamefont {Schreier}}, \bibinfo
  {author} {\bibfnamefont {J.}~\bibnamefont {Lotze}}, \bibinfo {author}
  {\bibfnamefont {M.}~\bibnamefont {Pernpeintner}}, \bibinfo {author}
  {\bibfnamefont {S.}~\bibnamefont {Meyer}}, \bibinfo {author} {\bibfnamefont
  {H.}~\bibnamefont {Huebl}}, \bibinfo {author} {\bibfnamefont
  {R.}~\bibnamefont {Gross}}, \bibinfo {author} {\bibfnamefont
  {A.}~\bibnamefont {Kamra}}, \bibinfo {author} {\bibfnamefont
  {J.}~\bibnamefont {Xiao}},  \emph {et~al.},\ }\bibfield  {title} {\bibinfo
  {title} {Experimental test of the spin mixing interface conductivity
  concept},\ }\href@noop {} {\bibfield  {journal} {\bibinfo  {journal}
  {Physical review letters}\ }\textbf {\bibinfo {volume} {111}},\ \bibinfo
  {pages} {176601} (\bibinfo {year} {2013})}\BibitemShut {NoStop}%
\bibitem [{\citenamefont {Ulloa}\ \emph {et~al.}(2019)\citenamefont {Ulloa},
  \citenamefont {Tomadin}, \citenamefont {Shan}, \citenamefont {Polini},
  \citenamefont {Van~Wees},\ and\ \citenamefont {Duine}}]{ulloa2019nonlocal}%
  \BibitemOpen
  \bibfield  {author} {\bibinfo {author} {\bibfnamefont {C.}~\bibnamefont
  {Ulloa}}, \bibinfo {author} {\bibfnamefont {A.}~\bibnamefont {Tomadin}},
  \bibinfo {author} {\bibfnamefont {J.}~\bibnamefont {Shan}}, \bibinfo {author}
  {\bibfnamefont {M.}~\bibnamefont {Polini}}, \bibinfo {author} {\bibfnamefont
  {B.}~\bibnamefont {Van~Wees}}, \ and\ \bibinfo {author} {\bibfnamefont
  {R.}~\bibnamefont {Duine}},\ }\bibfield  {title} {\bibinfo {title} {Nonlocal
  spin transport as a probe of viscous magnon fluids},\ }\href@noop {}
  {\bibfield  {journal} {\bibinfo  {journal} {Physical Review Letters}\
  }\textbf {\bibinfo {volume} {123}},\ \bibinfo {pages} {117203} (\bibinfo
  {year} {2019})}\BibitemShut {NoStop}%
\bibitem [{\citenamefont {Bender}\ \emph {et~al.}(2012)\citenamefont {Bender},
  \citenamefont {Duine},\ and\ \citenamefont
  {Tserkovnyak}}]{bender2012electronic}%
  \BibitemOpen
  \bibfield  {author} {\bibinfo {author} {\bibfnamefont {S.~A.}\ \bibnamefont
  {Bender}}, \bibinfo {author} {\bibfnamefont {R.~A.}\ \bibnamefont {Duine}}, \
  and\ \bibinfo {author} {\bibfnamefont {Y.}~\bibnamefont {Tserkovnyak}},\
  }\bibfield  {title} {\bibinfo {title} {Electronic pumping of quasiequilibrium
  bose-einstein-condensed magnons},\ }\href@noop {} {\bibfield  {journal}
  {\bibinfo  {journal} {Physical review letters}\ }\textbf {\bibinfo {volume}
  {108}},\ \bibinfo {pages} {246601} (\bibinfo {year} {2012})}\BibitemShut
  {NoStop}%
\bibitem [{\citenamefont {Tserkovnyak}\ \emph {et~al.}(2016)\citenamefont
  {Tserkovnyak}, \citenamefont {Bender}, \citenamefont {Duine},\ and\
  \citenamefont {Flebus}}]{tserkovnyak2016bose}%
  \BibitemOpen
  \bibfield  {author} {\bibinfo {author} {\bibfnamefont {Y.}~\bibnamefont
  {Tserkovnyak}}, \bibinfo {author} {\bibfnamefont {S.~A.}\ \bibnamefont
  {Bender}}, \bibinfo {author} {\bibfnamefont {R.~A.}\ \bibnamefont {Duine}}, \
  and\ \bibinfo {author} {\bibfnamefont {B.}~\bibnamefont {Flebus}},\
  }\bibfield  {title} {\bibinfo {title} {Bose-einstein condensation of magnons
  pumped by the bulk spin seebeck effect},\ }\href@noop {} {\bibfield
  {journal} {\bibinfo  {journal} {Physical Review B}\ }\textbf {\bibinfo
  {volume} {93}},\ \bibinfo {pages} {100402} (\bibinfo {year}
  {2016})}\BibitemShut {NoStop}%
\bibitem [{\citenamefont {Safranski}\ \emph {et~al.}(2017)\citenamefont
  {Safranski}, \citenamefont {Barsukov}, \citenamefont {Lee}, \citenamefont
  {Schneider}, \citenamefont {Jara}, \citenamefont {Smith}, \citenamefont
  {Chang}, \citenamefont {Lenz}, \citenamefont {Lindner}, \citenamefont
  {Tserkovnyak} \emph {et~al.}}]{safranski2017spin}%
  \BibitemOpen
  \bibfield  {author} {\bibinfo {author} {\bibfnamefont {C.}~\bibnamefont
  {Safranski}}, \bibinfo {author} {\bibfnamefont {I.}~\bibnamefont {Barsukov}},
  \bibinfo {author} {\bibfnamefont {H.~K.}\ \bibnamefont {Lee}}, \bibinfo
  {author} {\bibfnamefont {T.}~\bibnamefont {Schneider}}, \bibinfo {author}
  {\bibfnamefont {A.}~\bibnamefont {Jara}}, \bibinfo {author} {\bibfnamefont
  {A.}~\bibnamefont {Smith}}, \bibinfo {author} {\bibfnamefont
  {H.}~\bibnamefont {Chang}}, \bibinfo {author} {\bibfnamefont
  {K.}~\bibnamefont {Lenz}}, \bibinfo {author} {\bibfnamefont {J.}~\bibnamefont
  {Lindner}}, \bibinfo {author} {\bibfnamefont {Y.}~\bibnamefont
  {Tserkovnyak}},  \emph {et~al.},\ }\bibfield  {title} {\bibinfo {title} {Spin
  caloritronic nano-oscillator},\ }\href@noop {} {\bibfield  {journal}
  {\bibinfo  {journal} {Nature communications}\ }\textbf {\bibinfo {volume}
  {8}},\ \bibinfo {pages} {1} (\bibinfo {year} {2017})}\BibitemShut {NoStop}%
\bibitem [{\citenamefont {Thiery}\ \emph
  {et~al.}(2018{\natexlab{a}})\citenamefont {Thiery}, \citenamefont {Draveny},
  \citenamefont {Naletov}, \citenamefont {Vila}, \citenamefont {Attan{\'e}},
  \citenamefont {Beign{\'e}}, \citenamefont {de~Loubens}, \citenamefont
  {Viret}, \citenamefont {Beaulieu}, \citenamefont {Youssef} \emph
  {et~al.}}]{thiery2018nonlinear}%
  \BibitemOpen
  \bibfield  {author} {\bibinfo {author} {\bibfnamefont {N.}~\bibnamefont
  {Thiery}}, \bibinfo {author} {\bibfnamefont {A.}~\bibnamefont {Draveny}},
  \bibinfo {author} {\bibfnamefont {V.}~\bibnamefont {Naletov}}, \bibinfo
  {author} {\bibfnamefont {L.}~\bibnamefont {Vila}}, \bibinfo {author}
  {\bibfnamefont {J.}~\bibnamefont {Attan{\'e}}}, \bibinfo {author}
  {\bibfnamefont {C.}~\bibnamefont {Beign{\'e}}}, \bibinfo {author}
  {\bibfnamefont {G.}~\bibnamefont {de~Loubens}}, \bibinfo {author}
  {\bibfnamefont {M.}~\bibnamefont {Viret}}, \bibinfo {author} {\bibfnamefont
  {N.}~\bibnamefont {Beaulieu}}, \bibinfo {author} {\bibfnamefont {J.~B.}\
  \bibnamefont {Youssef}},  \emph {et~al.},\ }\bibfield  {title} {\bibinfo
  {title} {Nonlinear spin conductance of yttrium iron garnet thin films driven
  by large spin-orbit torque},\ }\href@noop {} {\bibfield  {journal} {\bibinfo
  {journal} {Physical Review B}\ }\textbf {\bibinfo {volume} {97}},\ \bibinfo
  {pages} {060409} (\bibinfo {year} {2018}{\natexlab{a}})}\BibitemShut
  {NoStop}%
\bibitem [{\citenamefont {Schneider}\ \emph {et~al.}(2020)\citenamefont
  {Schneider}, \citenamefont {Br{\"a}cher}, \citenamefont {Breitbach},
  \citenamefont {Lauer}, \citenamefont {Pirro}, \citenamefont {Bozhko},
  \citenamefont {Musiienko-Shmarova}, \citenamefont {Heinz}, \citenamefont
  {Wang}, \citenamefont {Meyer} \emph {et~al.}}]{schneider2020bose}%
  \BibitemOpen
  \bibfield  {author} {\bibinfo {author} {\bibfnamefont {M.}~\bibnamefont
  {Schneider}}, \bibinfo {author} {\bibfnamefont {T.}~\bibnamefont
  {Br{\"a}cher}}, \bibinfo {author} {\bibfnamefont {D.}~\bibnamefont
  {Breitbach}}, \bibinfo {author} {\bibfnamefont {V.}~\bibnamefont {Lauer}},
  \bibinfo {author} {\bibfnamefont {P.}~\bibnamefont {Pirro}}, \bibinfo
  {author} {\bibfnamefont {D.~A.}\ \bibnamefont {Bozhko}}, \bibinfo {author}
  {\bibfnamefont {H.~Y.}\ \bibnamefont {Musiienko-Shmarova}}, \bibinfo {author}
  {\bibfnamefont {B.}~\bibnamefont {Heinz}}, \bibinfo {author} {\bibfnamefont
  {Q.}~\bibnamefont {Wang}}, \bibinfo {author} {\bibfnamefont {T.}~\bibnamefont
  {Meyer}},  \emph {et~al.},\ }\bibfield  {title} {\bibinfo {title}
  {Bose--einstein condensation of quasiparticles by rapid cooling},\
  }\href@noop {} {\bibfield  {journal} {\bibinfo  {journal} {Nature
  Nanotechnology}\ ,\ \bibinfo {pages} {1}} (\bibinfo {year}
  {2020})}\BibitemShut {NoStop}%
\bibitem [{\citenamefont {Weiler}\ \emph {et~al.}(2012)\citenamefont {Weiler},
  \citenamefont {Althammer}, \citenamefont {Czeschka}, \citenamefont {Huebl},
  \citenamefont {Wagner}, \citenamefont {Opel}, \citenamefont {Imort},
  \citenamefont {Reiss}, \citenamefont {Thomas}, \citenamefont {Gross} \emph
  {et~al.}}]{weiler2012local}%
  \BibitemOpen
  \bibfield  {author} {\bibinfo {author} {\bibfnamefont {M.}~\bibnamefont
  {Weiler}}, \bibinfo {author} {\bibfnamefont {M.}~\bibnamefont {Althammer}},
  \bibinfo {author} {\bibfnamefont {F.~D.}\ \bibnamefont {Czeschka}}, \bibinfo
  {author} {\bibfnamefont {H.}~\bibnamefont {Huebl}}, \bibinfo {author}
  {\bibfnamefont {M.~S.}\ \bibnamefont {Wagner}}, \bibinfo {author}
  {\bibfnamefont {M.}~\bibnamefont {Opel}}, \bibinfo {author} {\bibfnamefont
  {I.-M.}\ \bibnamefont {Imort}}, \bibinfo {author} {\bibfnamefont
  {G.}~\bibnamefont {Reiss}}, \bibinfo {author} {\bibfnamefont
  {A.}~\bibnamefont {Thomas}}, \bibinfo {author} {\bibfnamefont
  {R.}~\bibnamefont {Gross}},  \emph {et~al.},\ }\bibfield  {title} {\bibinfo
  {title} {Local charge and spin currents in magnetothermal landscapes},\
  }\href@noop {} {\bibfield  {journal} {\bibinfo  {journal} {Physical review
  letters}\ }\textbf {\bibinfo {volume} {108}},\ \bibinfo {pages} {106602}
  (\bibinfo {year} {2012})}\BibitemShut {NoStop}%
\bibitem [{\citenamefont {Jamison}\ \emph {et~al.}(2019)\citenamefont
  {Jamison}, \citenamefont {Yang}, \citenamefont {Giles}, \citenamefont
  {Brangham}, \citenamefont {Wu}, \citenamefont {Hammel}, \citenamefont
  {Yang},\ and\ \citenamefont {Myers}}]{jamison2019long}%
  \BibitemOpen
  \bibfield  {author} {\bibinfo {author} {\bibfnamefont {J.~S.}\ \bibnamefont
  {Jamison}}, \bibinfo {author} {\bibfnamefont {Z.}~\bibnamefont {Yang}},
  \bibinfo {author} {\bibfnamefont {B.~L.}\ \bibnamefont {Giles}}, \bibinfo
  {author} {\bibfnamefont {J.~T.}\ \bibnamefont {Brangham}}, \bibinfo {author}
  {\bibfnamefont {G.}~\bibnamefont {Wu}}, \bibinfo {author} {\bibfnamefont
  {P.~C.}\ \bibnamefont {Hammel}}, \bibinfo {author} {\bibfnamefont
  {F.}~\bibnamefont {Yang}}, \ and\ \bibinfo {author} {\bibfnamefont {R.~C.}\
  \bibnamefont {Myers}},\ }\bibfield  {title} {\bibinfo {title} {Long lifetime
  of thermally excited magnons in bulk yttrium iron garnet},\ }\href@noop {}
  {\bibfield  {journal} {\bibinfo  {journal} {Physical Review B}\ }\textbf
  {\bibinfo {volume} {100}},\ \bibinfo {pages} {134402} (\bibinfo {year}
  {2019})}\BibitemShut {NoStop}%
\bibitem [{\citenamefont {Olsson}\ \emph {et~al.}(2020)\citenamefont {Olsson},
  \citenamefont {An}, \citenamefont {Fiete}, \citenamefont {Zhou},
  \citenamefont {Shi},\ and\ \citenamefont {Li}}]{olsson2020pure}%
  \BibitemOpen
  \bibfield  {author} {\bibinfo {author} {\bibfnamefont {K.~S.}\ \bibnamefont
  {Olsson}}, \bibinfo {author} {\bibfnamefont {K.}~\bibnamefont {An}}, \bibinfo
  {author} {\bibfnamefont {G.~A.}\ \bibnamefont {Fiete}}, \bibinfo {author}
  {\bibfnamefont {J.}~\bibnamefont {Zhou}}, \bibinfo {author} {\bibfnamefont
  {L.}~\bibnamefont {Shi}}, \ and\ \bibinfo {author} {\bibfnamefont
  {X.}~\bibnamefont {Li}},\ }\bibfield  {title} {\bibinfo {title} {Pure spin
  current and magnon chemical potential in a nonequilibrium magnetic
  insulator},\ }\href {\doibase 10.1103/PhysRevX.10.021029} {\bibfield
  {journal} {\bibinfo  {journal} {Phys. Rev. X}\ }\textbf {\bibinfo {volume}
  {10}},\ \bibinfo {pages} {021029} (\bibinfo {year} {2020})}\BibitemShut
  {NoStop}%
\bibitem [{\citenamefont {Cornelissen}\ \emph {et~al.}(2015)\citenamefont
  {Cornelissen}, \citenamefont {Liu}, \citenamefont {Duine}, \citenamefont
  {Youssef},\ and\ \citenamefont {Van~Wees}}]{cornelissen2015long}%
  \BibitemOpen
  \bibfield  {author} {\bibinfo {author} {\bibfnamefont {L.}~\bibnamefont
  {Cornelissen}}, \bibinfo {author} {\bibfnamefont {J.}~\bibnamefont {Liu}},
  \bibinfo {author} {\bibfnamefont {R.}~\bibnamefont {Duine}}, \bibinfo
  {author} {\bibfnamefont {J.~B.}\ \bibnamefont {Youssef}}, \ and\ \bibinfo
  {author} {\bibfnamefont {B.}~\bibnamefont {Van~Wees}},\ }\bibfield  {title}
  {\bibinfo {title} {Long-distance transport of magnon spin information in a
  magnetic insulator at room temperature},\ }\href@noop {} {\bibfield
  {journal} {\bibinfo  {journal} {Nature Physics}\ }\textbf {\bibinfo {volume}
  {11}},\ \bibinfo {pages} {1022} (\bibinfo {year} {2015})}\BibitemShut
  {NoStop}%
\bibitem [{\citenamefont {Giles}\ \emph {et~al.}(2017)\citenamefont {Giles},
  \citenamefont {Yang}, \citenamefont {Jamison}, \citenamefont {Gomez-Perez},
  \citenamefont {V{\'e}lez}, \citenamefont {Hueso}, \citenamefont {Casanova},\
  and\ \citenamefont {Myers}}]{giles2017thermally}%
  \BibitemOpen
  \bibfield  {author} {\bibinfo {author} {\bibfnamefont {B.~L.}\ \bibnamefont
  {Giles}}, \bibinfo {author} {\bibfnamefont {Z.}~\bibnamefont {Yang}},
  \bibinfo {author} {\bibfnamefont {J.~S.}\ \bibnamefont {Jamison}}, \bibinfo
  {author} {\bibfnamefont {J.~M.}\ \bibnamefont {Gomez-Perez}}, \bibinfo
  {author} {\bibfnamefont {S.}~\bibnamefont {V{\'e}lez}}, \bibinfo {author}
  {\bibfnamefont {L.~E.}\ \bibnamefont {Hueso}}, \bibinfo {author}
  {\bibfnamefont {F.}~\bibnamefont {Casanova}}, \ and\ \bibinfo {author}
  {\bibfnamefont {R.~C.}\ \bibnamefont {Myers}},\ }\bibfield  {title} {\bibinfo
  {title} {Thermally driven long-range magnon spin currents in yttrium iron
  garnet due to intrinsic spin seebeck effect},\ }\href@noop {} {\bibfield
  {journal} {\bibinfo  {journal} {Physical Review B}\ }\textbf {\bibinfo
  {volume} {96}},\ \bibinfo {pages} {180412} (\bibinfo {year}
  {2017})}\BibitemShut {NoStop}%
\bibitem [{\citenamefont {Boona}\ and\ \citenamefont
  {Heremans}(2014)}]{boona2014magnon}%
  \BibitemOpen
  \bibfield  {author} {\bibinfo {author} {\bibfnamefont {S.~R.}\ \bibnamefont
  {Boona}}\ and\ \bibinfo {author} {\bibfnamefont {J.~P.}\ \bibnamefont
  {Heremans}},\ }\bibfield  {title} {\bibinfo {title} {Magnon thermal mean free
  path in yttrium iron garnet},\ }\href@noop {} {\bibfield  {journal} {\bibinfo
   {journal} {Physical Review B}\ }\textbf {\bibinfo {volume} {90}},\ \bibinfo
  {pages} {064421} (\bibinfo {year} {2014})}\BibitemShut {NoStop}%
\bibitem [{\citenamefont {Dyson}(1956)}]{Dyson1956}%
  \BibitemOpen
  \bibfield  {author} {\bibinfo {author} {\bibfnamefont {F.~J.}\ \bibnamefont
  {Dyson}},\ }\bibfield  {title} {\bibinfo {title} {General theory of spin-wave
  interactions},\ }\href {\doibase 10.1103/physrev.102.1217} {\bibfield
  {journal} {\bibinfo  {journal} {Physical Review}\ }\textbf {\bibinfo {volume}
  {102}},\ \bibinfo {pages} {1217} (\bibinfo {year} {1956})}\BibitemShut
  {NoStop}%
\bibitem [{\citenamefont {Flebus}\ \emph {et~al.}(2016)\citenamefont {Flebus},
  \citenamefont {Bender}, \citenamefont {Tserkovnyak},\ and\ \citenamefont
  {Duine}}]{flebus2016two}%
  \BibitemOpen
  \bibfield  {author} {\bibinfo {author} {\bibfnamefont {B.}~\bibnamefont
  {Flebus}}, \bibinfo {author} {\bibfnamefont {S.}~\bibnamefont {Bender}},
  \bibinfo {author} {\bibfnamefont {Y.}~\bibnamefont {Tserkovnyak}}, \ and\
  \bibinfo {author} {\bibfnamefont {R.}~\bibnamefont {Duine}},\ }\bibfield
  {title} {\bibinfo {title} {Two-fluid theory for spin superfluidity in
  magnetic insulators},\ }\href@noop {} {\bibfield  {journal} {\bibinfo
  {journal} {Physical review letters}\ }\textbf {\bibinfo {volume} {116}},\
  \bibinfo {pages} {117201} (\bibinfo {year} {2016})}\BibitemShut {NoStop}%
\bibitem [{Note1()}]{Note1}%
  \BibitemOpen
  \bibinfo {note} {$\alpha \sim 10^{-4}$ is a lower bound for the Gilbert
  damping in YIG thin films with thickness $<$~100nm}\BibitemShut {NoStop}%
\bibitem [{\citenamefont {Cherepanov}\ \emph {et~al.}(1993)\citenamefont
  {Cherepanov}, \citenamefont {Kolokolov},\ and\ \citenamefont
  {L'vov}}]{cherepanov1993saga}%
  \BibitemOpen
  \bibfield  {author} {\bibinfo {author} {\bibfnamefont {V.}~\bibnamefont
  {Cherepanov}}, \bibinfo {author} {\bibfnamefont {I.}~\bibnamefont
  {Kolokolov}}, \ and\ \bibinfo {author} {\bibfnamefont {V.}~\bibnamefont
  {L'vov}},\ }\bibfield  {title} {\bibinfo {title} {The saga of yig: spectra,
  thermodynamics, interaction and relaxation of magnons in a complex magnet},\
  }\href@noop {} {\bibfield  {journal} {\bibinfo  {journal} {Physics reports}\
  }\textbf {\bibinfo {volume} {229}},\ \bibinfo {pages} {81} (\bibinfo {year}
  {1993})}\BibitemShut {NoStop}%
\bibitem [{\citenamefont {Plant}(1977)}]{plant1977spinwave}%
  \BibitemOpen
  \bibfield  {author} {\bibinfo {author} {\bibfnamefont {J.}~\bibnamefont
  {Plant}},\ }\bibfield  {title} {\bibinfo {title} {Spinwave dispersion curves
  for yttrium iron garnet},\ }\href@noop {} {\bibfield  {journal} {\bibinfo
  {journal} {Journal of Physics C: Solid State Physics}\ }\textbf {\bibinfo
  {volume} {10}},\ \bibinfo {pages} {4805} (\bibinfo {year}
  {1977})}\BibitemShut {NoStop}%
\bibitem [{\citenamefont {Barker}\ and\ \citenamefont
  {Bauer}(2016)}]{barker2016thermal}%
  \BibitemOpen
  \bibfield  {author} {\bibinfo {author} {\bibfnamefont {J.}~\bibnamefont
  {Barker}}\ and\ \bibinfo {author} {\bibfnamefont {G.~E.}\ \bibnamefont
  {Bauer}},\ }\bibfield  {title} {\bibinfo {title} {Thermal spin dynamics of
  yttrium iron garnet},\ }\href@noop {} {\bibfield  {journal} {\bibinfo
  {journal} {Physical review letters}\ }\textbf {\bibinfo {volume} {117}},\
  \bibinfo {pages} {217201} (\bibinfo {year} {2016})}\BibitemShut {NoStop}%
\bibitem [{\citenamefont {Kehlberger}\ \emph {et~al.}(2015)\citenamefont
  {Kehlberger}, \citenamefont {Ritzmann}, \citenamefont {Hinzke}, \citenamefont
  {Guo}, \citenamefont {Cramer}, \citenamefont {Jakob}, \citenamefont
  {Onbasli}, \citenamefont {Kim}, \citenamefont {Ross}, \citenamefont
  {Jungfleisch} \emph {et~al.}}]{kehlberger2015length}%
  \BibitemOpen
  \bibfield  {author} {\bibinfo {author} {\bibfnamefont {A.}~\bibnamefont
  {Kehlberger}}, \bibinfo {author} {\bibfnamefont {U.}~\bibnamefont
  {Ritzmann}}, \bibinfo {author} {\bibfnamefont {D.}~\bibnamefont {Hinzke}},
  \bibinfo {author} {\bibfnamefont {E.-J.}\ \bibnamefont {Guo}}, \bibinfo
  {author} {\bibfnamefont {J.}~\bibnamefont {Cramer}}, \bibinfo {author}
  {\bibfnamefont {G.}~\bibnamefont {Jakob}}, \bibinfo {author} {\bibfnamefont
  {M.~C.}\ \bibnamefont {Onbasli}}, \bibinfo {author} {\bibfnamefont {D.~H.}\
  \bibnamefont {Kim}}, \bibinfo {author} {\bibfnamefont {C.~A.}\ \bibnamefont
  {Ross}}, \bibinfo {author} {\bibfnamefont {M.~B.}\ \bibnamefont
  {Jungfleisch}},  \emph {et~al.},\ }\bibfield  {title} {\bibinfo {title}
  {Length scale of the spin seebeck effect},\ }\href@noop {} {\bibfield
  {journal} {\bibinfo  {journal} {Physical review letters}\ }\textbf {\bibinfo
  {volume} {115}},\ \bibinfo {pages} {096602} (\bibinfo {year}
  {2015})}\BibitemShut {NoStop}%
\bibitem [{\citenamefont {Prakash}\ \emph {et~al.}(2018)\citenamefont
  {Prakash}, \citenamefont {Flebus}, \citenamefont {Brangham}, \citenamefont
  {Yang}, \citenamefont {Tserkovnyak},\ and\ \citenamefont
  {Heremans}}]{prakash2018evidence}%
  \BibitemOpen
  \bibfield  {author} {\bibinfo {author} {\bibfnamefont {A.}~\bibnamefont
  {Prakash}}, \bibinfo {author} {\bibfnamefont {B.}~\bibnamefont {Flebus}},
  \bibinfo {author} {\bibfnamefont {J.}~\bibnamefont {Brangham}}, \bibinfo
  {author} {\bibfnamefont {F.}~\bibnamefont {Yang}}, \bibinfo {author}
  {\bibfnamefont {Y.}~\bibnamefont {Tserkovnyak}}, \ and\ \bibinfo {author}
  {\bibfnamefont {J.~P.}\ \bibnamefont {Heremans}},\ }\bibfield  {title}
  {\bibinfo {title} {Evidence for the role of the magnon energy relaxation
  length in the spin seebeck effect},\ }\href@noop {} {\bibfield  {journal}
  {\bibinfo  {journal} {Physical Review B}\ }\textbf {\bibinfo {volume} {97}},\
  \bibinfo {pages} {020408} (\bibinfo {year} {2018})}\BibitemShut {NoStop}%
\bibitem [{\citenamefont {Agrawal}\ \emph {et~al.}(2014)\citenamefont
  {Agrawal}, \citenamefont {Vasyuchka}, \citenamefont {Serga}, \citenamefont
  {Kirihara}, \citenamefont {Pirro}, \citenamefont {Langner}, \citenamefont
  {Jungfleisch}, \citenamefont {Chumak}, \citenamefont {Papaioannou},\ and\
  \citenamefont {Hillebrands}}]{agrawal2014role}%
  \BibitemOpen
  \bibfield  {author} {\bibinfo {author} {\bibfnamefont {M.}~\bibnamefont
  {Agrawal}}, \bibinfo {author} {\bibfnamefont {V.}~\bibnamefont {Vasyuchka}},
  \bibinfo {author} {\bibfnamefont {A.}~\bibnamefont {Serga}}, \bibinfo
  {author} {\bibfnamefont {A.}~\bibnamefont {Kirihara}}, \bibinfo {author}
  {\bibfnamefont {P.}~\bibnamefont {Pirro}}, \bibinfo {author} {\bibfnamefont
  {T.}~\bibnamefont {Langner}}, \bibinfo {author} {\bibfnamefont
  {M.}~\bibnamefont {Jungfleisch}}, \bibinfo {author} {\bibfnamefont
  {A.}~\bibnamefont {Chumak}}, \bibinfo {author} {\bibfnamefont {E.~T.}\
  \bibnamefont {Papaioannou}}, \ and\ \bibinfo {author} {\bibfnamefont
  {B.}~\bibnamefont {Hillebrands}},\ }\bibfield  {title} {\bibinfo {title}
  {Role of bulk-magnon transport in the temporal evolution of the longitudinal
  spin-seebeck effect},\ }\href@noop {} {\bibfield  {journal} {\bibinfo
  {journal} {Physical Review B}\ }\textbf {\bibinfo {volume} {89}},\ \bibinfo
  {pages} {224414} (\bibinfo {year} {2014})}\BibitemShut {NoStop}%
\bibitem [{\citenamefont {Shan}\ \emph {et~al.}(2016)\citenamefont {Shan},
  \citenamefont {Cornelissen}, \citenamefont {Vlietstra}, \citenamefont
  {Youssef}, \citenamefont {Kuschel}, \citenamefont {Duine},\ and\
  \citenamefont {Van~Wees}}]{shan2016influence}%
  \BibitemOpen
  \bibfield  {author} {\bibinfo {author} {\bibfnamefont {J.}~\bibnamefont
  {Shan}}, \bibinfo {author} {\bibfnamefont {L.~J.}\ \bibnamefont
  {Cornelissen}}, \bibinfo {author} {\bibfnamefont {N.}~\bibnamefont
  {Vlietstra}}, \bibinfo {author} {\bibfnamefont {J.~B.}\ \bibnamefont
  {Youssef}}, \bibinfo {author} {\bibfnamefont {T.}~\bibnamefont {Kuschel}},
  \bibinfo {author} {\bibfnamefont {R.~A.}\ \bibnamefont {Duine}}, \ and\
  \bibinfo {author} {\bibfnamefont {B.~J.}\ \bibnamefont {Van~Wees}},\
  }\bibfield  {title} {\bibinfo {title} {Influence of yttrium iron garnet
  thickness and heater opacity on the nonlocal transport of electrically and
  thermally excited magnons},\ }\href@noop {} {\bibfield  {journal} {\bibinfo
  {journal} {Physical Review B}\ }\textbf {\bibinfo {volume} {94}},\ \bibinfo
  {pages} {174437} (\bibinfo {year} {2016})}\BibitemShut {NoStop}%
\bibitem [{\citenamefont {Adachi}\ \emph {et~al.}(2013)\citenamefont {Adachi},
  \citenamefont {Uchida}, \citenamefont {Saitoh},\ and\ \citenamefont
  {Maekawa}}]{adachi2013theory}%
  \BibitemOpen
  \bibfield  {author} {\bibinfo {author} {\bibfnamefont {H.}~\bibnamefont
  {Adachi}}, \bibinfo {author} {\bibfnamefont {K.-i.}\ \bibnamefont {Uchida}},
  \bibinfo {author} {\bibfnamefont {E.}~\bibnamefont {Saitoh}}, \ and\ \bibinfo
  {author} {\bibfnamefont {S.}~\bibnamefont {Maekawa}},\ }\bibfield  {title}
  {\bibinfo {title} {Theory of the spin seebeck effect},\ }\href@noop {}
  {\bibfield  {journal} {\bibinfo  {journal} {Reports on Progress in Physics}\
  }\textbf {\bibinfo {volume} {76}},\ \bibinfo {pages} {036501} (\bibinfo
  {year} {2013})}\BibitemShut {NoStop}%
\bibitem [{\citenamefont {Cornelissen}\ \emph {et~al.}(2016)\citenamefont
  {Cornelissen}, \citenamefont {Peters}, \citenamefont {Bauer}, \citenamefont
  {Duine},\ and\ \citenamefont {van Wees}}]{cornelissen2016magnon}%
  \BibitemOpen
  \bibfield  {author} {\bibinfo {author} {\bibfnamefont {L.~J.}\ \bibnamefont
  {Cornelissen}}, \bibinfo {author} {\bibfnamefont {K.~J.}\ \bibnamefont
  {Peters}}, \bibinfo {author} {\bibfnamefont {G.~E.}\ \bibnamefont {Bauer}},
  \bibinfo {author} {\bibfnamefont {R.}~\bibnamefont {Duine}}, \ and\ \bibinfo
  {author} {\bibfnamefont {B.~J.}\ \bibnamefont {van Wees}},\ }\bibfield
  {title} {\bibinfo {title} {Magnon spin transport driven by the magnon
  chemical potential in a magnetic insulator},\ }\href@noop {} {\bibfield
  {journal} {\bibinfo  {journal} {Physical Review B}\ }\textbf {\bibinfo
  {volume} {94}},\ \bibinfo {pages} {014412} (\bibinfo {year}
  {2016})}\BibitemShut {NoStop}%
\bibitem [{\citenamefont {Beaulieu}\ \emph {et~al.}(2018)\citenamefont
  {Beaulieu}, \citenamefont {Kervarec}, \citenamefont {Thiery}, \citenamefont
  {Klein}, \citenamefont {Naletov}, \citenamefont {Hurdequint}, \citenamefont
  {de~Loubens}, \citenamefont {Youssef},\ and\ \citenamefont
  {Vukadinovic}}]{beaulieu2018temperature}%
  \BibitemOpen
  \bibfield  {author} {\bibinfo {author} {\bibfnamefont {N.}~\bibnamefont
  {Beaulieu}}, \bibinfo {author} {\bibfnamefont {N.}~\bibnamefont {Kervarec}},
  \bibinfo {author} {\bibfnamefont {N.}~\bibnamefont {Thiery}}, \bibinfo
  {author} {\bibfnamefont {O.}~\bibnamefont {Klein}}, \bibinfo {author}
  {\bibfnamefont {V.}~\bibnamefont {Naletov}}, \bibinfo {author} {\bibfnamefont
  {H.}~\bibnamefont {Hurdequint}}, \bibinfo {author} {\bibfnamefont
  {G.}~\bibnamefont {de~Loubens}}, \bibinfo {author} {\bibfnamefont {J.~B.}\
  \bibnamefont {Youssef}}, \ and\ \bibinfo {author} {\bibfnamefont
  {N.}~\bibnamefont {Vukadinovic}},\ }\bibfield  {title} {\bibinfo {title}
  {Temperature dependence of magnetic properties of a ultrathin yttrium-iron
  garnet film grown by liquid phase epitaxy: Effect of a pt overlayer},\
  }\href@noop {} {\bibfield  {journal} {\bibinfo  {journal} {IEEE Magnetics
  Letters}\ }\textbf {\bibinfo {volume} {9}},\ \bibinfo {pages} {1} (\bibinfo
  {year} {2018})}\BibitemShut {NoStop}%
\bibitem [{sup()}]{suppl}%
  \BibitemOpen
  \href@noop {} {\bibinfo  {journal} {See Supplemental Material at [URL will be
  inserted by publisher] for details of simulation and the effect of
  interfacial contribution}\ }\BibitemShut {NoStop}%
\bibitem [{\citenamefont {Thiery}\ \emph
  {et~al.}(2018{\natexlab{b}})\citenamefont {Thiery}, \citenamefont {Naletov},
  \citenamefont {Vila}, \citenamefont {Marty}, \citenamefont {Brenac},
  \citenamefont {Jacquot}, \citenamefont {de~Loubens}, \citenamefont {Viret},
  \citenamefont {Anane}, \citenamefont {Cros} \emph
  {et~al.}}]{thiery2018electrical}%
  \BibitemOpen
\bibfield  {journal} {  }\bibfield  {author} {\bibinfo {author} {\bibfnamefont
  {N.}~\bibnamefont {Thiery}}, \bibinfo {author} {\bibfnamefont
  {V.}~\bibnamefont {Naletov}}, \bibinfo {author} {\bibfnamefont
  {L.}~\bibnamefont {Vila}}, \bibinfo {author} {\bibfnamefont {A.}~\bibnamefont
  {Marty}}, \bibinfo {author} {\bibfnamefont {A.}~\bibnamefont {Brenac}},
  \bibinfo {author} {\bibfnamefont {J.-F.}\ \bibnamefont {Jacquot}}, \bibinfo
  {author} {\bibfnamefont {G.}~\bibnamefont {de~Loubens}}, \bibinfo {author}
  {\bibfnamefont {M.}~\bibnamefont {Viret}}, \bibinfo {author} {\bibfnamefont
  {A.}~\bibnamefont {Anane}}, \bibinfo {author} {\bibfnamefont
  {V.}~\bibnamefont {Cros}},  \emph {et~al.},\ }\bibfield  {title} {\bibinfo
  {title} {Electrical properties of epitaxial yttrium iron garnet ultrathin
  films at high temperatures},\ }\href@noop {} {\bibfield  {journal} {\bibinfo
  {journal} {Physical Review B}\ }\textbf {\bibinfo {volume} {97}},\ \bibinfo
  {pages} {064422} (\bibinfo {year} {2018}{\natexlab{b}})}\BibitemShut
  {NoStop}%
\bibitem [{\citenamefont {Chen}\ \emph {et~al.}(2013)\citenamefont {Chen},
  \citenamefont {Takahashi}, \citenamefont {Nakayama}, \citenamefont
  {Althammer}, \citenamefont {Goennenwein}, \citenamefont {Saitoh},\ and\
  \citenamefont {Bauer}}]{chen2013theory}%
  \BibitemOpen
  \bibfield  {author} {\bibinfo {author} {\bibfnamefont {Y.-T.}\ \bibnamefont
  {Chen}}, \bibinfo {author} {\bibfnamefont {S.}~\bibnamefont {Takahashi}},
  \bibinfo {author} {\bibfnamefont {H.}~\bibnamefont {Nakayama}}, \bibinfo
  {author} {\bibfnamefont {M.}~\bibnamefont {Althammer}}, \bibinfo {author}
  {\bibfnamefont {S.~T.}\ \bibnamefont {Goennenwein}}, \bibinfo {author}
  {\bibfnamefont {E.}~\bibnamefont {Saitoh}}, \ and\ \bibinfo {author}
  {\bibfnamefont {G.~E.}\ \bibnamefont {Bauer}},\ }\bibfield  {title} {\bibinfo
  {title} {Theory of spin hall magnetoresistance},\ }\href@noop {} {\bibfield
  {journal} {\bibinfo  {journal} {Physical Review B}\ }\textbf {\bibinfo
  {volume} {87}},\ \bibinfo {pages} {144411} (\bibinfo {year}
  {2013})}\BibitemShut {NoStop}%
\bibitem [{\citenamefont {Nakayama}\ \emph {et~al.}(2013)\citenamefont
  {Nakayama}, \citenamefont {Althammer}, \citenamefont {Chen}, \citenamefont
  {Uchida}, \citenamefont {Kajiwara}, \citenamefont {Kikuchi}, \citenamefont
  {Ohtani}, \citenamefont {Gepr{\"a}gs}, \citenamefont {Opel}, \citenamefont
  {Takahashi} \emph {et~al.}}]{nakayama2013spin}%
  \BibitemOpen
  \bibfield  {author} {\bibinfo {author} {\bibfnamefont {H.}~\bibnamefont
  {Nakayama}}, \bibinfo {author} {\bibfnamefont {M.}~\bibnamefont {Althammer}},
  \bibinfo {author} {\bibfnamefont {Y.-T.}\ \bibnamefont {Chen}}, \bibinfo
  {author} {\bibfnamefont {K.}~\bibnamefont {Uchida}}, \bibinfo {author}
  {\bibfnamefont {Y.}~\bibnamefont {Kajiwara}}, \bibinfo {author}
  {\bibfnamefont {D.}~\bibnamefont {Kikuchi}}, \bibinfo {author} {\bibfnamefont
  {T.}~\bibnamefont {Ohtani}}, \bibinfo {author} {\bibfnamefont
  {S.}~\bibnamefont {Gepr{\"a}gs}}, \bibinfo {author} {\bibfnamefont
  {M.}~\bibnamefont {Opel}}, \bibinfo {author} {\bibfnamefont {S.}~\bibnamefont
  {Takahashi}},  \emph {et~al.},\ }\bibfield  {title} {\bibinfo {title} {Spin
  hall magnetoresistance induced by a nonequilibrium proximity effect},\
  }\href@noop {} {\bibfield  {journal} {\bibinfo  {journal} {Physical review
  letters}\ }\textbf {\bibinfo {volume} {110}},\ \bibinfo {pages} {206601}
  (\bibinfo {year} {2013})}\BibitemShut {NoStop}%
\bibitem [{\citenamefont {Hahn}\ \emph {et~al.}(2013)\citenamefont {Hahn},
  \citenamefont {De~Loubens}, \citenamefont {Klein}, \citenamefont {Viret},
  \citenamefont {Naletov},\ and\ \citenamefont
  {Youssef}}]{hahn2013comparative}%
  \BibitemOpen
  \bibfield  {author} {\bibinfo {author} {\bibfnamefont {C.}~\bibnamefont
  {Hahn}}, \bibinfo {author} {\bibfnamefont {G.}~\bibnamefont {De~Loubens}},
  \bibinfo {author} {\bibfnamefont {O.}~\bibnamefont {Klein}}, \bibinfo
  {author} {\bibfnamefont {M.}~\bibnamefont {Viret}}, \bibinfo {author}
  {\bibfnamefont {V.~V.}\ \bibnamefont {Naletov}}, \ and\ \bibinfo {author}
  {\bibfnamefont {J.~B.}\ \bibnamefont {Youssef}},\ }\bibfield  {title}
  {\bibinfo {title} {Comparative measurements of inverse spin hall effects and
  magnetoresistance in yig/pt and yig/ta},\ }\href@noop {} {\bibfield
  {journal} {\bibinfo  {journal} {Physical Review B}\ }\textbf {\bibinfo
  {volume} {87}},\ \bibinfo {pages} {174417} (\bibinfo {year}
  {2013})}\BibitemShut {NoStop}%
\bibitem [{\citenamefont {Uchida}\ \emph {et~al.}(2008)\citenamefont {Uchida},
  \citenamefont {Takahashi}, \citenamefont {Harii}, \citenamefont {Ieda},
  \citenamefont {Koshibae}, \citenamefont {Ando}, \citenamefont {Maekawa},\
  and\ \citenamefont {Saitoh}}]{uchida2008observation}%
  \BibitemOpen
  \bibfield  {author} {\bibinfo {author} {\bibfnamefont {K.}~\bibnamefont
  {Uchida}}, \bibinfo {author} {\bibfnamefont {S.}~\bibnamefont {Takahashi}},
  \bibinfo {author} {\bibfnamefont {K.}~\bibnamefont {Harii}}, \bibinfo
  {author} {\bibfnamefont {J.}~\bibnamefont {Ieda}}, \bibinfo {author}
  {\bibfnamefont {W.}~\bibnamefont {Koshibae}}, \bibinfo {author}
  {\bibfnamefont {K.}~\bibnamefont {Ando}}, \bibinfo {author} {\bibfnamefont
  {S.}~\bibnamefont {Maekawa}}, \ and\ \bibinfo {author} {\bibfnamefont
  {E.}~\bibnamefont {Saitoh}},\ }\bibfield  {title} {\bibinfo {title}
  {Observation of the spin seebeck effect},\ }\href@noop {} {\bibfield
  {journal} {\bibinfo  {journal} {Nature}\ }\textbf {\bibinfo {volume} {455}},\
  \bibinfo {pages} {778} (\bibinfo {year} {2008})}\BibitemShut {NoStop}%
\bibitem [{\citenamefont {Gilleo}\ and\ \citenamefont
  {Geller}(1958)}]{gilleo1958magnetic}%
  \BibitemOpen
  \bibfield  {author} {\bibinfo {author} {\bibfnamefont {M.}~\bibnamefont
  {Gilleo}}\ and\ \bibinfo {author} {\bibfnamefont {S.}~\bibnamefont
  {Geller}},\ }\bibfield  {title} {\bibinfo {title} {Magnetic and
  crystallographic properties of substituted yttrium-iron garnet, $3\textsc{Y}
  _2\textsc{O}_3 \cdot x \textsc{M}_2\textsc{O}_ 3 \cdot(5-
  x)\textsc{F}e_2\textsc{O}_3$},\ }\href@noop {} {\bibfield  {journal}
  {\bibinfo  {journal} {Physical Review}\ }\textbf {\bibinfo {volume} {110}},\
  \bibinfo {pages} {73} (\bibinfo {year} {1958})}\BibitemShut {NoStop}%
\bibitem [{\citenamefont {Guo}\ \emph {et~al.}(2016)\citenamefont {Guo},
  \citenamefont {Cramer}, \citenamefont {Kehlberger}, \citenamefont {Ferguson},
  \citenamefont {MacLaren}, \citenamefont {Jakob},\ and\ \citenamefont
  {Kl{\"a}ui}}]{guo2016influence}%
  \BibitemOpen
  \bibfield  {author} {\bibinfo {author} {\bibfnamefont {E.-J.}\ \bibnamefont
  {Guo}}, \bibinfo {author} {\bibfnamefont {J.}~\bibnamefont {Cramer}},
  \bibinfo {author} {\bibfnamefont {A.}~\bibnamefont {Kehlberger}}, \bibinfo
  {author} {\bibfnamefont {C.~A.}\ \bibnamefont {Ferguson}}, \bibinfo {author}
  {\bibfnamefont {D.~A.}\ \bibnamefont {MacLaren}}, \bibinfo {author}
  {\bibfnamefont {G.}~\bibnamefont {Jakob}}, \ and\ \bibinfo {author}
  {\bibfnamefont {M.}~\bibnamefont {Kl{\"a}ui}},\ }\bibfield  {title} {\bibinfo
  {title} {Influence of thickness and interface on the low-temperature
  enhancement of the spin seebeck effect in yig films},\ }\href@noop {}
  {\bibfield  {journal} {\bibinfo  {journal} {Physical Review X}\ }\textbf
  {\bibinfo {volume} {6}},\ \bibinfo {pages} {031012} (\bibinfo {year}
  {2016})}\BibitemShut {NoStop}%
\bibitem [{\citenamefont {Ganzhorn}\ \emph {et~al.}(2017)\citenamefont
  {Ganzhorn}, \citenamefont {Wimmer}, \citenamefont {Cramer}, \citenamefont
  {Schlitz}, \citenamefont {Gepr{\"a}gs}, \citenamefont {Jakob}, \citenamefont
  {Gross}, \citenamefont {Huebl}, \citenamefont {Kl{\"a}ui},\ and\
  \citenamefont {Goennenwein}}]{ganzhorn2017temperature}%
  \BibitemOpen
  \bibfield  {author} {\bibinfo {author} {\bibfnamefont {K.}~\bibnamefont
  {Ganzhorn}}, \bibinfo {author} {\bibfnamefont {T.}~\bibnamefont {Wimmer}},
  \bibinfo {author} {\bibfnamefont {J.}~\bibnamefont {Cramer}}, \bibinfo
  {author} {\bibfnamefont {R.}~\bibnamefont {Schlitz}}, \bibinfo {author}
  {\bibfnamefont {S.}~\bibnamefont {Gepr{\"a}gs}}, \bibinfo {author}
  {\bibfnamefont {G.}~\bibnamefont {Jakob}}, \bibinfo {author} {\bibfnamefont
  {R.}~\bibnamefont {Gross}}, \bibinfo {author} {\bibfnamefont
  {H.}~\bibnamefont {Huebl}}, \bibinfo {author} {\bibfnamefont
  {M.}~\bibnamefont {Kl{\"a}ui}}, \ and\ \bibinfo {author} {\bibfnamefont
  {S.~T.}\ \bibnamefont {Goennenwein}},\ }\bibfield  {title} {\bibinfo {title}
  {Temperature dependence of the non-local spin seebeck effect in yig/pt
  nanostructures},\ }\href@noop {} {\bibfield  {journal} {\bibinfo  {journal}
  {AIP Advances}\ }\textbf {\bibinfo {volume} {7}},\ \bibinfo {pages} {085102}
  (\bibinfo {year} {2017})}\BibitemShut {NoStop}%
\bibitem [{\citenamefont {Zhou}\ \emph {et~al.}(2017)\citenamefont {Zhou},
  \citenamefont {Shi}, \citenamefont {Han}, \citenamefont {Yang}, \citenamefont
  {Rao}, \citenamefont {Zhang}, \citenamefont {Lang}, \citenamefont {Zhou},
  \citenamefont {Pan},\ and\ \citenamefont {Song}}]{zhou2017lateral}%
  \BibitemOpen
  \bibfield  {author} {\bibinfo {author} {\bibfnamefont {X.}~\bibnamefont
  {Zhou}}, \bibinfo {author} {\bibfnamefont {G.}~\bibnamefont {Shi}}, \bibinfo
  {author} {\bibfnamefont {J.}~\bibnamefont {Han}}, \bibinfo {author}
  {\bibfnamefont {Q.}~\bibnamefont {Yang}}, \bibinfo {author} {\bibfnamefont
  {Y.}~\bibnamefont {Rao}}, \bibinfo {author} {\bibfnamefont {H.}~\bibnamefont
  {Zhang}}, \bibinfo {author} {\bibfnamefont {L.}~\bibnamefont {Lang}},
  \bibinfo {author} {\bibfnamefont {S.}~\bibnamefont {Zhou}}, \bibinfo {author}
  {\bibfnamefont {F.}~\bibnamefont {Pan}}, \ and\ \bibinfo {author}
  {\bibfnamefont {C.}~\bibnamefont {Song}},\ }\bibfield  {title} {\bibinfo
  {title} {Lateral transport properties of thermally excited magnons in yttrium
  iron garnet films},\ }\href@noop {} {\bibfield  {journal} {\bibinfo
  {journal} {Applied Physics Letters}\ }\textbf {\bibinfo {volume} {110}},\
  \bibinfo {pages} {062407} (\bibinfo {year} {2017})}\BibitemShut {NoStop}%
\bibitem [{\citenamefont {Castel}\ \emph {et~al.}(2012)\citenamefont {Castel},
  \citenamefont {Vlietstra}, \citenamefont {Ben~Youssef},\ and\ \citenamefont
  {van Wees}}]{castel2012platinum}%
  \BibitemOpen
  \bibfield  {author} {\bibinfo {author} {\bibfnamefont {V.}~\bibnamefont
  {Castel}}, \bibinfo {author} {\bibfnamefont {N.}~\bibnamefont {Vlietstra}},
  \bibinfo {author} {\bibfnamefont {J.}~\bibnamefont {Ben~Youssef}}, \ and\
  \bibinfo {author} {\bibfnamefont {B.~J.}\ \bibnamefont {van Wees}},\
  }\bibfield  {title} {\bibinfo {title} {Platinum thickness dependence of the
  inverse spin-hall voltage from spin pumping in a hybrid yttrium iron
  garnet/platinum system},\ }\href@noop {} {\bibfield  {journal} {\bibinfo
  {journal} {Applied Physics Letters}\ }\textbf {\bibinfo {volume} {101}},\
  \bibinfo {pages} {132414} (\bibinfo {year} {2012})}\BibitemShut {NoStop}%
\bibitem [{\citenamefont {Xiao}\ \emph {et~al.}(2010)\citenamefont {Xiao},
  \citenamefont {Bauer}, \citenamefont {Uchida}, \citenamefont {Saitoh},
  \citenamefont {Maekawa} \emph {et~al.}}]{xiao2010theory}%
  \BibitemOpen
  \bibfield  {author} {\bibinfo {author} {\bibfnamefont {J.}~\bibnamefont
  {Xiao}}, \bibinfo {author} {\bibfnamefont {G.~E.}\ \bibnamefont {Bauer}},
  \bibinfo {author} {\bibfnamefont {K.-c.}\ \bibnamefont {Uchida}}, \bibinfo
  {author} {\bibfnamefont {E.}~\bibnamefont {Saitoh}}, \bibinfo {author}
  {\bibfnamefont {S.}~\bibnamefont {Maekawa}},  \emph {et~al.},\ }\bibfield
  {title} {\bibinfo {title} {Theory of magnon-driven spin seebeck effect},\
  }\href@noop {} {\bibfield  {journal} {\bibinfo  {journal} {Physical Review
  B}\ }\textbf {\bibinfo {volume} {81}},\ \bibinfo {pages} {214418} (\bibinfo
  {year} {2010})}\BibitemShut {NoStop}%
\bibitem [{\citenamefont {Bender}\ and\ \citenamefont
  {Tserkovnyak}(2015)}]{bender2015interfacial}%
  \BibitemOpen
  \bibfield  {author} {\bibinfo {author} {\bibfnamefont {S.~A.}\ \bibnamefont
  {Bender}}\ and\ \bibinfo {author} {\bibfnamefont {Y.}~\bibnamefont
  {Tserkovnyak}},\ }\bibfield  {title} {\bibinfo {title} {Interfacial spin and
  heat transfer between metals and magnetic insulators},\ }\href@noop {}
  {\bibfield  {journal} {\bibinfo  {journal} {Physical Review B}\ }\textbf
  {\bibinfo {volume} {91}},\ \bibinfo {pages} {140402} (\bibinfo {year}
  {2015})}\BibitemShut {NoStop}%
\bibitem [{Note2()}]{Note2}%
  \BibitemOpen
  \bibinfo {note} {Our pulse method, which allows the injection of large
  current densities in the Pt, does not have the dynamical range to follow the
  SSE signal at distances above 10~$\mu $m, where the signal is diminished by
  three orders of magnitude. This prevents us from studying if there is another
  tail at this exponential decay with a characteristic length scale of about
  tens of micron as reported in other studies.}\BibitemShut {Stop}%
\bibitem [{Note3()}]{Note3}%
  \BibitemOpen
  \bibinfo {note} {Probably the energy relaxation length of magnons is also a
  multiscale issue}\BibitemShut {NoStop}%
\end{thebibliography}%


\begin{thebibliography}{10}%
\makeatletter
\providecommand \@ifxundefined [1]{%
 \@ifx{#1\undefined}
}%
\providecommand \@ifnum [1]{%
 \ifnum #1\expandafter \@firstoftwo
 \else \expandafter \@secondoftwo
 \fi
}%
\providecommand \@ifx [1]{%
 \ifx #1\expandafter \@firstoftwo
 \else \expandafter \@secondoftwo
 \fi
}%
\providecommand \natexlab [1]{#1}%
\providecommand \enquote  [1]{``#1''}%
\providecommand \bibnamefont  [1]{#1}%
\providecommand \bibfnamefont [1]{#1}%
\providecommand \citenamefont [1]{#1}%
\providecommand \href@noop [0]{\@secondoftwo}%
\providecommand \href [0]{\begingroup \@sanitize@url \@href}%
\providecommand \@href[1]{\@@startlink{#1}\@@href}%
\providecommand \@@href[1]{\endgroup#1\@@endlink}%
\providecommand \@sanitize@url [0]{\catcode `\\12\catcode `\$12\catcode
  `\&12\catcode `\#12\catcode `\^12\catcode `\_12\catcode `\%12\relax}%
\providecommand \@@startlink[1]{}%
\providecommand \@@endlink[0]{}%
\providecommand \url  [0]{\begingroup\@sanitize@url \@url }%
\providecommand \@url [1]{\endgroup\@href {#1}{\urlprefix }}%
\providecommand \urlprefix  [0]{URL }%
\providecommand \Eprint [0]{\href }%
\providecommand \doibase [0]{http://dx.doi.org/}%
\providecommand \selectlanguage [0]{\@gobble}%
\providecommand \bibinfo  [0]{\@secondoftwo}%
\providecommand \bibfield  [0]{\@secondoftwo}%
\providecommand \translation [1]{[#1]}%
\providecommand \BibitemOpen [0]{}%
\providecommand \bibitemStop [0]{}%
\providecommand \bibitemNoStop [0]{.\EOS\space}%
\providecommand \EOS [0]{\spacefactor3000\relax}%
\providecommand \BibitemShut  [1]{\csname bibitem#1\endcsname}%
\let\auto@bib@innerbib\@empty
\bibitem [{\citenamefont {Thiery}\ \emph {et~al.}(2018)\citenamefont {Thiery},
  \citenamefont {Draveny}, \citenamefont {Naletov}, \citenamefont {Vila},
  \citenamefont {Attan{\'e}}, \citenamefont {Beign{\'e}}, \citenamefont
  {de~Loubens}, \citenamefont {Viret}, \citenamefont {Beaulieu}, \citenamefont
  {Youssef} \emph {et~al.}}]{thiery2018nonlinear}%
  \BibitemOpen
  \bibfield  {author} {\bibinfo {author} {\bibfnamefont {N.}~\bibnamefont
  {Thiery}}, \bibinfo {author} {\bibfnamefont {A.}~\bibnamefont {Draveny}},
  \bibinfo {author} {\bibfnamefont {V.}~\bibnamefont {Naletov}}, \bibinfo
  {author} {\bibfnamefont {L.}~\bibnamefont {Vila}}, \bibinfo {author}
  {\bibfnamefont {J.}~\bibnamefont {Attan{\'e}}}, \bibinfo {author}
  {\bibfnamefont {C.}~\bibnamefont {Beign{\'e}}}, \bibinfo {author}
  {\bibfnamefont {G.}~\bibnamefont {de~Loubens}}, \bibinfo {author}
  {\bibfnamefont {M.}~\bibnamefont {Viret}}, \bibinfo {author} {\bibfnamefont
  {N.}~\bibnamefont {Beaulieu}}, \bibinfo {author} {\bibfnamefont {J.~B.}\
  \bibnamefont {Youssef}},  \emph {et~al.},\ }\bibfield  {title} {\bibinfo
  {title} {Nonlinear spin conductance of yttrium iron garnet thin films driven
  by large spin-orbit torque},\ }\href@noop {} {\bibfield  {journal} {\bibinfo
  {journal} {Physical Review B}\ }\textbf {\bibinfo {volume} {97}},\ \bibinfo
  {pages} {060409} (\bibinfo {year} {2018})}\BibitemShut {NoStop}%
\bibitem [{\citenamefont {Slack}\ and\ \citenamefont
  {Oliver}(1971)}]{slack1971thermal}%
  \BibitemOpen
  \bibfield  {author} {\bibinfo {author} {\bibfnamefont {G.~A.}\ \bibnamefont
  {Slack}}\ and\ \bibinfo {author} {\bibfnamefont {D.}~\bibnamefont {Oliver}},\
  }\bibfield  {title} {\bibinfo {title} {Thermal conductivity of garnets and
  phonon scattering by rare-earth ions},\ }\href@noop {} {\bibfield  {journal}
  {\bibinfo  {journal} {Physical Review B}\ }\textbf {\bibinfo {volume} {4}},\
  \bibinfo {pages} {592} (\bibinfo {year} {1971})}\BibitemShut {NoStop}%
\bibitem [{\citenamefont {Zhang}\ \emph {et~al.}(2005)\citenamefont {Zhang},
  \citenamefont {Xie}, \citenamefont {Fujii}, \citenamefont {Ago},
  \citenamefont {Takahashi}, \citenamefont {Ikuta}, \citenamefont {Abe},\ and\
  \citenamefont {Shimizu}}]{zhang2005thermal}%
  \BibitemOpen
  \bibfield  {author} {\bibinfo {author} {\bibfnamefont {X.}~\bibnamefont
  {Zhang}}, \bibinfo {author} {\bibfnamefont {H.}~\bibnamefont {Xie}}, \bibinfo
  {author} {\bibfnamefont {M.}~\bibnamefont {Fujii}}, \bibinfo {author}
  {\bibfnamefont {H.}~\bibnamefont {Ago}}, \bibinfo {author} {\bibfnamefont
  {K.}~\bibnamefont {Takahashi}}, \bibinfo {author} {\bibfnamefont
  {T.}~\bibnamefont {Ikuta}}, \bibinfo {author} {\bibfnamefont
  {H.}~\bibnamefont {Abe}}, \ and\ \bibinfo {author} {\bibfnamefont
  {T.}~\bibnamefont {Shimizu}},\ }\bibfield  {title} {\bibinfo {title} {Thermal
  and electrical conductivity of a suspended platinum nanofilm},\ }\href@noop
  {} {\bibfield  {journal} {\bibinfo  {journal} {Applied Physics Letters}\
  }\textbf {\bibinfo {volume} {86}},\ \bibinfo {pages} {171912} (\bibinfo
  {year} {2005})}\BibitemShut {NoStop}%
\bibitem [{\citenamefont {Volkov}\ \emph {et~al.}(1976)\citenamefont {Volkov},
  \citenamefont {Palatnik},\ and\ \citenamefont
  {Pugachev}}]{volkov1976investigation}%
  \BibitemOpen
  \bibfield  {author} {\bibinfo {author} {\bibfnamefont {Y.~A.}\ \bibnamefont
  {Volkov}}, \bibinfo {author} {\bibfnamefont {L.}~\bibnamefont {Palatnik}}, \
  and\ \bibinfo {author} {\bibfnamefont {A.}~\bibnamefont {Pugachev}},\
  }\bibfield  {title} {\bibinfo {title} {Investigation of the thermal
  properties of thin aluminum films},\ }\href@noop {} {\bibfield  {journal}
  {\bibinfo  {journal} {Zh. Eksp. Teor. Fiz}\ }\textbf {\bibinfo {volume}
  {70}},\ \bibinfo {pages} {2244} (\bibinfo {year} {1976})}\BibitemShut
  {NoStop}%
\bibitem [{\citenamefont {Lee}\ and\ \citenamefont
  {Cahill}(1997)}]{lee1997heat}%
  \BibitemOpen
  \bibfield  {author} {\bibinfo {author} {\bibfnamefont {S.-M.}\ \bibnamefont
  {Lee}}\ and\ \bibinfo {author} {\bibfnamefont {D.~G.}\ \bibnamefont
  {Cahill}},\ }\bibfield  {title} {\bibinfo {title} {Heat transport in thin
  dielectric films},\ }\href@noop {} {\bibfield  {journal} {\bibinfo  {journal}
  {Journal of applied physics}\ }\textbf {\bibinfo {volume} {81}},\ \bibinfo
  {pages} {2590} (\bibinfo {year} {1997})}\BibitemShut {NoStop}%
\bibitem [{\citenamefont {Uchida}\ \emph {et~al.}(2014)\citenamefont {Uchida},
  \citenamefont {Kikkawa}, \citenamefont {Miura}, \citenamefont {Shiomi},\ and\
  \citenamefont {Saitoh}}]{uchida2014quantitative}%
  \BibitemOpen
  \bibfield  {author} {\bibinfo {author} {\bibfnamefont {K.-i.}\ \bibnamefont
  {Uchida}}, \bibinfo {author} {\bibfnamefont {T.}~\bibnamefont {Kikkawa}},
  \bibinfo {author} {\bibfnamefont {A.}~\bibnamefont {Miura}}, \bibinfo
  {author} {\bibfnamefont {J.}~\bibnamefont {Shiomi}}, \ and\ \bibinfo {author}
  {\bibfnamefont {E.}~\bibnamefont {Saitoh}},\ }\bibfield  {title} {\bibinfo
  {title} {Quantitative temperature dependence of longitudinal spin seebeck
  effect at high temperatures},\ }\href@noop {} {\bibfield  {journal} {\bibinfo
   {journal} {Physical Review X}\ }\textbf {\bibinfo {volume} {4}},\ \bibinfo
  {pages} {041023} (\bibinfo {year} {2014})}\BibitemShut {NoStop}%
\bibitem [{\citenamefont {Beaulieu}\ \emph {et~al.}(2018)\citenamefont
  {Beaulieu}, \citenamefont {Kervarec}, \citenamefont {Thiery}, \citenamefont
  {Klein}, \citenamefont {Naletov}, \citenamefont {Hurdequint}, \citenamefont
  {de~Loubens}, \citenamefont {Youssef},\ and\ \citenamefont
  {Vukadinovic}}]{beaulieu2018temperature}%
  \BibitemOpen
  \bibfield  {author} {\bibinfo {author} {\bibfnamefont {N.}~\bibnamefont
  {Beaulieu}}, \bibinfo {author} {\bibfnamefont {N.}~\bibnamefont {Kervarec}},
  \bibinfo {author} {\bibfnamefont {N.}~\bibnamefont {Thiery}}, \bibinfo
  {author} {\bibfnamefont {O.}~\bibnamefont {Klein}}, \bibinfo {author}
  {\bibfnamefont {V.}~\bibnamefont {Naletov}}, \bibinfo {author} {\bibfnamefont
  {H.}~\bibnamefont {Hurdequint}}, \bibinfo {author} {\bibfnamefont
  {G.}~\bibnamefont {de~Loubens}}, \bibinfo {author} {\bibfnamefont {J.~B.}\
  \bibnamefont {Youssef}}, \ and\ \bibinfo {author} {\bibfnamefont
  {N.}~\bibnamefont {Vukadinovic}},\ }\bibfield  {title} {\bibinfo {title}
  {Temperature dependence of magnetic properties of a ultrathin yttrium-iron
  garnet film grown by liquid phase epitaxy: Effect of a pt overlayer},\
  }\href@noop {} {\bibfield  {journal} {\bibinfo  {journal} {IEEE Magnetics
  Letters}\ }\textbf {\bibinfo {volume} {9}},\ \bibinfo {pages} {1} (\bibinfo
  {year} {2018})}\BibitemShut {NoStop}%
\bibitem [{\citenamefont {Guo}\ \emph {et~al.}(2016)\citenamefont {Guo},
  \citenamefont {Cramer}, \citenamefont {Kehlberger}, \citenamefont {Ferguson},
  \citenamefont {MacLaren}, \citenamefont {Jakob},\ and\ \citenamefont
  {Kl{\"a}ui}}]{guo2016influence}%
  \BibitemOpen
  \bibfield  {author} {\bibinfo {author} {\bibfnamefont {E.-J.}\ \bibnamefont
  {Guo}}, \bibinfo {author} {\bibfnamefont {J.}~\bibnamefont {Cramer}},
  \bibinfo {author} {\bibfnamefont {A.}~\bibnamefont {Kehlberger}}, \bibinfo
  {author} {\bibfnamefont {C.~A.}\ \bibnamefont {Ferguson}}, \bibinfo {author}
  {\bibfnamefont {D.~A.}\ \bibnamefont {MacLaren}}, \bibinfo {author}
  {\bibfnamefont {G.}~\bibnamefont {Jakob}}, \ and\ \bibinfo {author}
  {\bibfnamefont {M.}~\bibnamefont {Kl{\"a}ui}},\ }\bibfield  {title} {\bibinfo
  {title} {Influence of thickness and interface on the low-temperature
  enhancement of the spin seebeck effect in yig films},\ }\href@noop {}
  {\bibfield  {journal} {\bibinfo  {journal} {Physical Review X}\ }\textbf
  {\bibinfo {volume} {6}},\ \bibinfo {pages} {031012} (\bibinfo {year}
  {2016})}\BibitemShut {NoStop}%
\bibitem [{\citenamefont {Cahill}\ \emph {et~al.}(2003)\citenamefont {Cahill},
  \citenamefont {Ford}, \citenamefont {Goodson}, \citenamefont {Mahan},
  \citenamefont {Majumdar}, \citenamefont {Maris}, \citenamefont {Merlin},\
  and\ \citenamefont {Phillpot}}]{cahill2003nanoscale}%
  \BibitemOpen
  \bibfield  {author} {\bibinfo {author} {\bibfnamefont {D.~G.}\ \bibnamefont
  {Cahill}}, \bibinfo {author} {\bibfnamefont {W.~K.}\ \bibnamefont {Ford}},
  \bibinfo {author} {\bibfnamefont {K.~E.}\ \bibnamefont {Goodson}}, \bibinfo
  {author} {\bibfnamefont {G.~D.}\ \bibnamefont {Mahan}}, \bibinfo {author}
  {\bibfnamefont {A.}~\bibnamefont {Majumdar}}, \bibinfo {author}
  {\bibfnamefont {H.~J.}\ \bibnamefont {Maris}}, \bibinfo {author}
  {\bibfnamefont {R.}~\bibnamefont {Merlin}}, \ and\ \bibinfo {author}
  {\bibfnamefont {S.~R.}\ \bibnamefont {Phillpot}},\ }\bibfield  {title}
  {\bibinfo {title} {Nanoscale thermal transport},\ }\href@noop {} {\bibfield
  {journal} {\bibinfo  {journal} {Journal of applied physics}\ }\textbf
  {\bibinfo {volume} {93}},\ \bibinfo {pages} {793} (\bibinfo {year}
  {2003})}\BibitemShut {NoStop}%
\bibitem [{\citenamefont {Bender}\ and\ \citenamefont
  {Tserkovnyak}(2015)}]{bender2015interfacial}%
  \BibitemOpen
  \bibfield  {author} {\bibinfo {author} {\bibfnamefont {S.~A.}\ \bibnamefont
  {Bender}}\ and\ \bibinfo {author} {\bibfnamefont {Y.}~\bibnamefont
  {Tserkovnyak}},\ }\bibfield  {title} {\bibinfo {title} {Interfacial spin and
  heat transfer between metals and magnetic insulators},\ }\href@noop {}
  {\bibfield  {journal} {\bibinfo  {journal} {Physical Review B}\ }\textbf
  {\bibinfo {volume} {91}},\ \bibinfo {pages} {140402} (\bibinfo {year}
  {2015})}\BibitemShut {NoStop}%
\end{thebibliography}%

\end{document}


\title{Supplemental material : Short-range Thermal Magnon Diffusion in Magnetic Garnets} 

\author{K. An}
\affiliation{Université Grenoble Alpes, CEA, CNRS, Grenoble INP, Spintec, 38054 Grenoble, France}

\author{R. Kohno}
\affiliation{Université Grenoble Alpes, CEA, CNRS, Grenoble INP, Spintec, 38054 Grenoble, France}

\author{N. Thiery}
\affiliation{Université Grenoble Alpes, CEA, CNRS, Grenoble INP, Spintec, 38054 Grenoble, France}

\author{D. Reitz}
\affiliation{Department of Physics and Astronomy, University of California, Los Angeles, California 90095, USA}

\author{L. Vila}
\affiliation{Université Grenoble Alpes, CEA, CNRS, Grenoble INP, Spintec, 38054 Grenoble, France}

\author{V. V. Naletov} 
\affiliation{Université Grenoble Alpes, CEA, CNRS, Grenoble INP, Spintec, 38054 Grenoble, France}
\affiliation{Institute of Physics, Kazan Federal University, Kazan
    420008, Russian Federation}

\author{N. Beaulieu} 
\affiliation{SPEC, CEA-Saclay, CNRS, Universit\'e Paris-Saclay,
  91191 Gif-sur-Yvette, France}
\affiliation{LabSTICC, CNRS, Universit\'e de Bretagne Occidentale,
  29238 Brest, France}

\author{J. Ben Youssef} 
\affiliation{LabSTICC, CNRS, Universit\'e de Bretagne Occidentale,
  29238 Brest, France}

\author{G. de Loubens} 
\affiliation{SPEC, CEA-Saclay, CNRS, Universit\'e Paris-Saclay,
  91191 Gif-sur-Yvette, France}

\author{Y. Tserkovnyak}
\affiliation{Department of Physics and Astronomy, University of California, Los Angeles, California 90095, USA}

\author{O. Klein}
\affiliation{Université Grenoble Alpes, CEA, CNRS, Grenoble INP, Spintec, 38054 Grenoble, France}

\date{\today}

\maketitle

\section{Temperature characterization}

\begin{figure}[b]
    \includegraphics[width=0.5\textwidth]{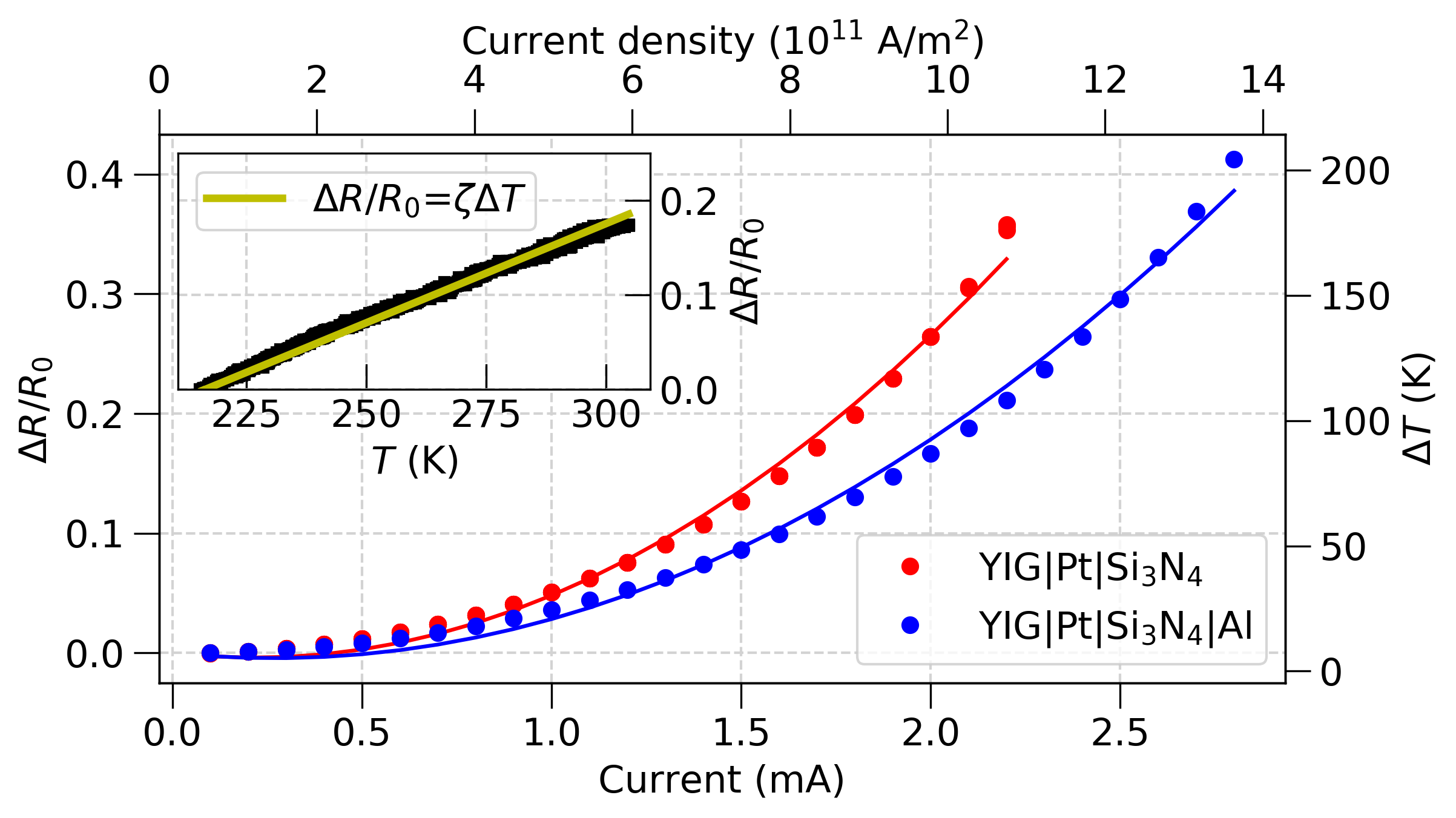}
    \caption{Resistance increase and the corresponding temperature elevation in the Pt strip as a function of the injected current with and without Al capping (red and blue dots, respectively). The solid red and blue lines are quadratic fits to the data. The inset shows the temperature dependence of Pt resistance. The yellow line is a linear fit to the data.
    }
    \label{FigS1}
\end{figure}

We characterize the temperature rise induced by Joule heating using the Pt resistance as a temperature sensor. The Pt injector is connected to a 6221 Keithley, which generates a 10$\,$ ms pulse current with a duty cylce of 10\%. The voltages are measured with a 2182A Keithley nano-voltmeter \cite{thiery2018nonlinear}. In the inset of Fig. \ref{FigS1} we plot the rational increase of Pt resistance as a function of ambient temperature between 220 and 300\,K. The change in resistance $\Delta R$ is linearly proportional to the temperature rise $\Delta T= T-T_0$, i.e., $\Delta R(T)$/$R_{0}$ = $\zeta\Delta T$, where $R_0$ is the initial resistance and $\zeta$ is the thermal coefficient of resistance and $T_0=300$~K is room temperature. We obtain $\zeta$ = $(2.1\pm 0.3) \times 10^{-3}$\,K$^{-1}$ for our Pt strip from the fit. The increase of resistance and corresponding temperature rise as a function of current are plotted in Fig. \ref{FigS1}. 
The Pt temperature increases quadratically with applying current owing to the
Joule heating ($\propto I^2$). The temperature rise in Pt is about 45 K lower
after the Al deposition at 2~mA (current density of $\sim 10^{12}$\,A/m$^{2}$). This indicates that the Al layer effectively spreads the heat from the Pt injector. 


\section{Details of theoretical calculation}
\begin{figure}[b]
    \includegraphics[width=0.5\textwidth]{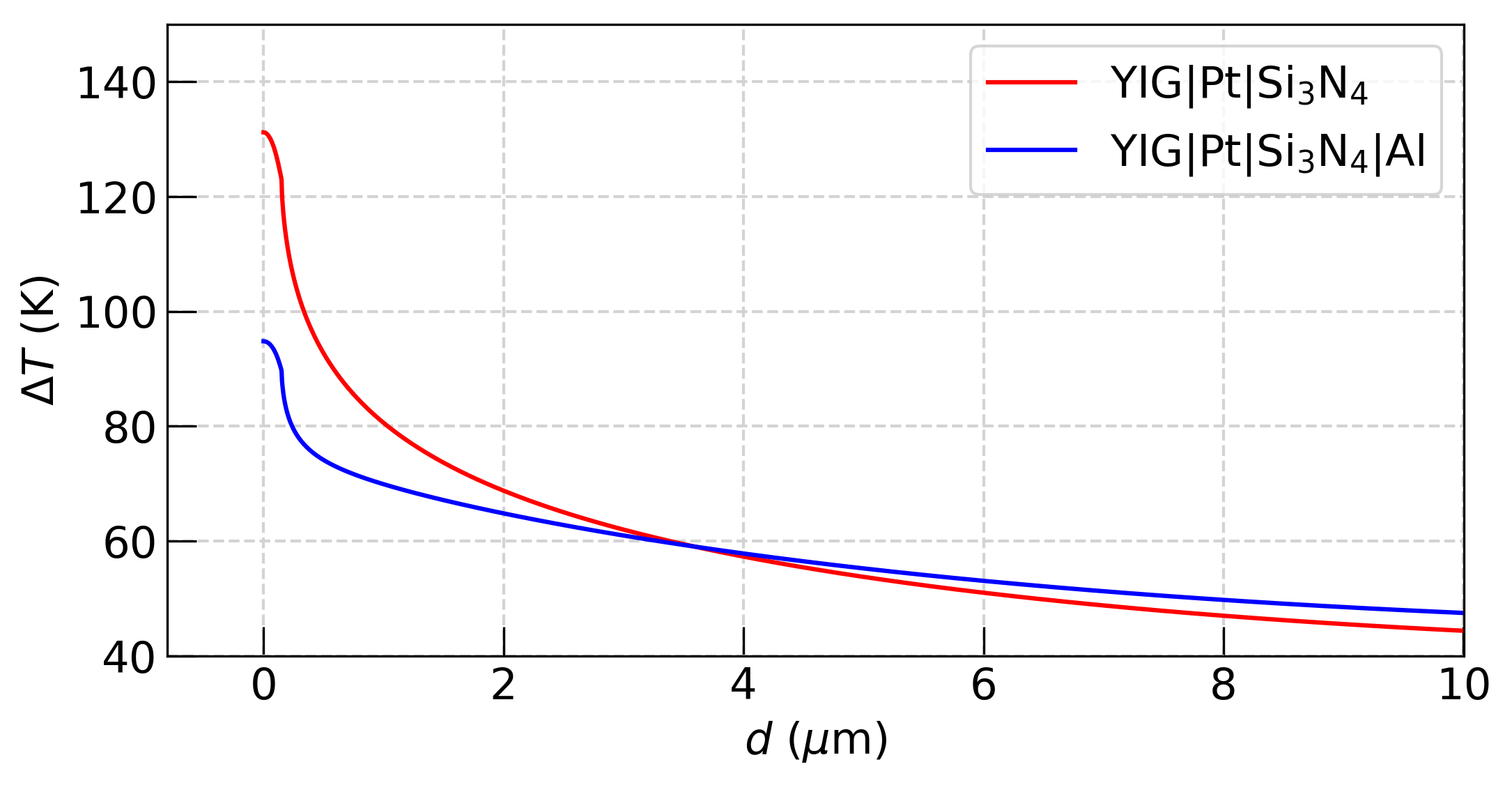}
    \caption{Calculated temperature rise profiles for two cases with and without the Al layer at 2 mA.}
    \label{FigS2}
\end{figure}

The temperature profiles and the chemcial potential are calculated in a 2D geometry using a finite element method, COMSOL. We choose a boundary condition that the top and side surfaces are thermally isolated and the bottom is held at room temperature. We assume that $\mu$ vanishes outside of YIG. The geometry was chosen to be the same as the actual sample size except that the lateral size of sample and the thickness of GGG are reduced to 30 $\mu$m to facilitate the calculation. The thermal conductivity parameters are 9, 7.4, 29, 220, and 0.5 Wm$^{-1}$K$^{-1}$ for GGG \cite{slack1971thermal}, YIG \cite{slack1971thermal}, Pt \cite{zhang2005thermal}, Al \cite{volkov1976investigation}, and Si$_3$N$_4$ \cite{lee1997heat}, respectively. The value of the spin Seebeck coefficient $\varsigma$ does not affect the decay profile and set to be 1\,$k_B$. The thermal magnon diffusion length $\lambda$ is varied. The calculated temperature profiles at the top of YIG surface are shown for the two cases in Fig. \ref{FigS2}.
The temperature difference at $d$ = 0 $\mu$m (the center of Pt injector strip) between two cases is 37 K, which roughly agrees with the measured temperature difference at 2 mA as shown in Fig. S1. To check the validity of calculated temperature profile, we measured the temperature rise at the position of the detector in case A. Our estimation yields a temperature drop of 46\% for the detector placed at $d$\,=\,0.5\,$\mu$m away from the injector. This is larger than the simulated temperature drop of 30\% over the same distance (red curve in \ref{FigS2}). The discrepancy may arise from (1) the simplification to 2D modeling, (2) possible difference in parameters between the simulation and measurement, and (3) heat taken away from the Pt detector, which is not accounted in the simulation. 


\section{Temperature dependence of the local SSE voltage}

\begin{figure}
    \includegraphics[width=0.5\textwidth]{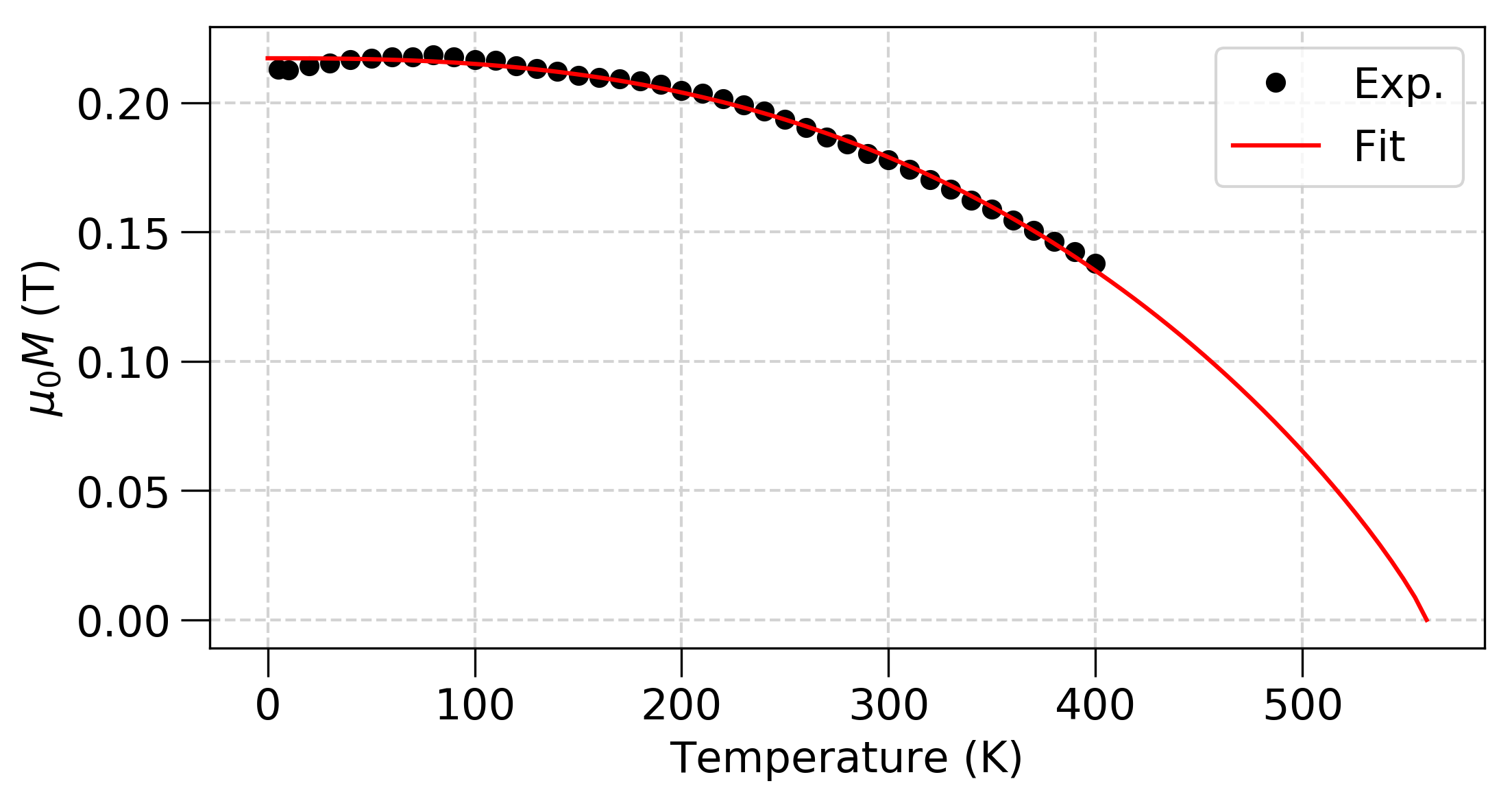}
    \caption{Measured temperature dependence of magnetization. The red solid line is a fit.}
    \label{FigS_MvsT}
\end{figure}

In Fig. 3(a) of the main text we fit the measured current dependence of $\Delta_1$ for case A. The measured local SSE voltage follows the analytical expression $\Delta_1 = S L_\text{Pt} \< \partial_z T \>$, where $S$ is the spin Seebeck coefficient, $L_\text{Pt}$ is the length of the Pt electrode, and $\<\partial_z T\>$ is the vertical temperature gradient across the YIG thickness. The latter is proportional to the temperature rise of the Pt injector: $ \<\partial_z T\> = (T-T_0)/ l_T$, where $T_0=300$~K is the substrate temperature and $l_T$ is the characteristic decay length of temperature from the top surface. By comparing the measured Pt temperature rise $T-T_0$ = 130 K at 2 mA with the expected  $\<\partial_z T\>=$ 10 K/56 nm as shown in Fig. 1(b) of the main text, we obtain $l_T \sim $ 730 nm. Assuming that the temperature dependence of $S$ is simply due to $\mu_0 M(T)$ (in contrast with the fitted temperature dependence used in a previous work \cite{uchida2014quantitative}), the expression for $\Delta_1$ becomes
\begin{equation}
    \Delta_1 (T) = C\dfrac{L_\text{Pt}}{l_T}  \left ( {T - T_0} \right ) \mu_0 M(T) ,
    \tag{S1}
    \label{eqS1}
\end{equation}
where $C \equiv S/(\mu_0 M(T))$. The temperature dependence of magnetization follows an empirical formula,
\begin{equation}
    \mu_0 M(T)=\mu_0  M_0 (1-(T/T_c)^{a})^{b},
    \tag{S2}
    \label{eqS2}
\end{equation}
where $\mu_0 M_0=0.217$~T is the YIG saturation magnetization at $T=0$\,K, while the exponents $a=2.6$ and $b=0.88$ are extracted from the fit as shown in Fig. \ref{FigS_MvsT} (data taken from \cite{beaulieu2018temperature}). The expression for $\Delta_1(T)$ is converted to $\Delta_1(I)$ using the temperature-current calibration curve in Fig. \ref{FigS1}. Then we use $C$ as a single fitting parameter to reproduce the observed behavior (solid red line in Fig. 3(a) of the main text). The fit yields $C$ = 0.45 $\mu$V\,K$^{-1}$\,T$^{-1}$. At room temperature, where the magnetization of YIG is about 0.178\,T, the spin Seebeck coefficient of our YIG$|$Pt system is about $S \approx$ 0.08\,$\mu$V\,K$^{-1}$, which agrees with a previous work \cite{guo2016influence}.



\section{Interfacial effect}

\begin{figure*}
    \includegraphics[width=1\textwidth]{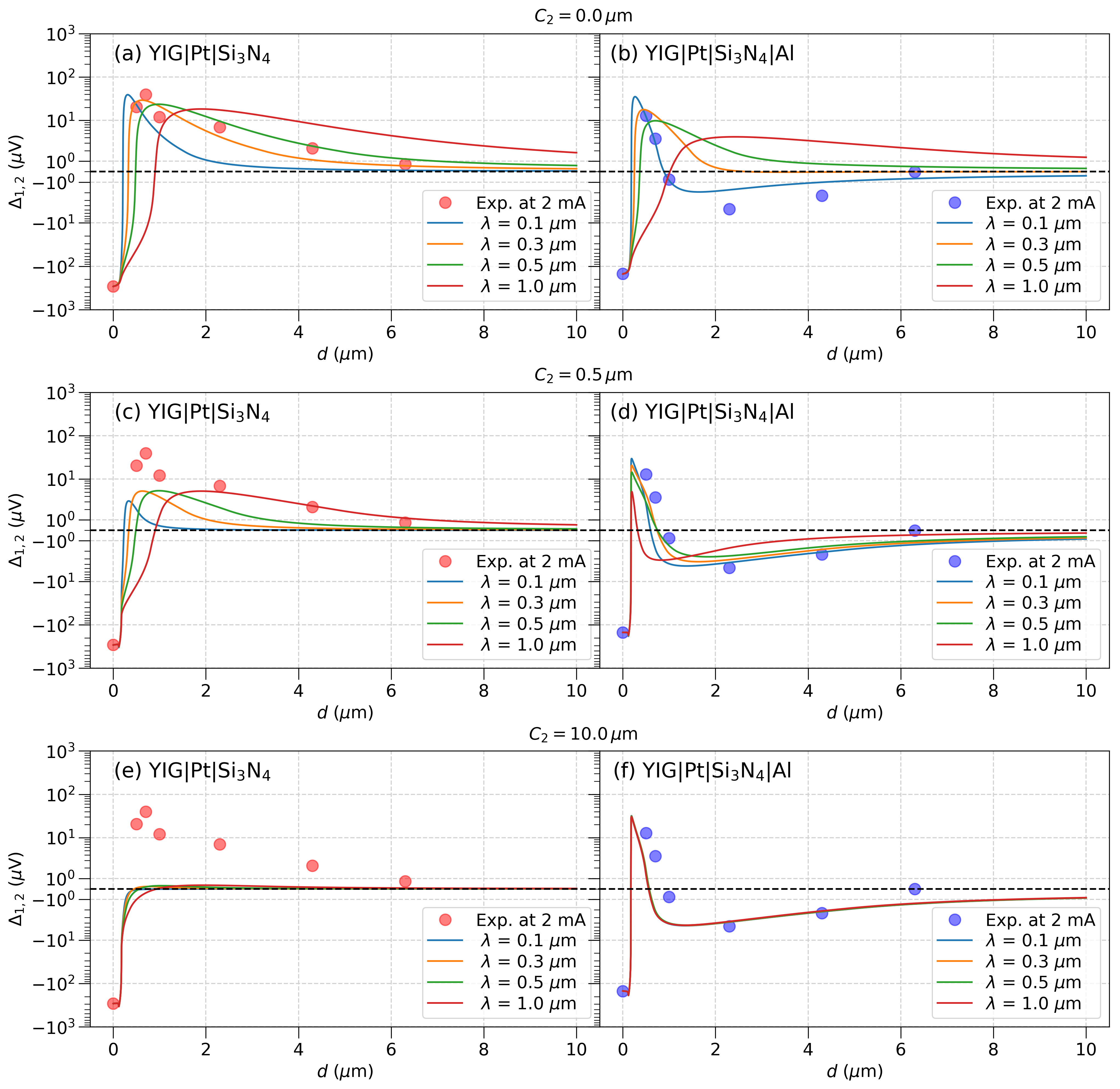}
    \caption{The simulated spatial profile of $\mu$ after normalization to the measured local $\Delta_1$. The simulation results are compared to the measured $\Delta$'s with varying $\lambda$ for (a,c,e) case A and (b,d,f) case B with $C_2$ = 0, 0.5, 10 $\mu$m. Only $\lambda$ = 100 and 300 nm show qualitative agreements for case B with $C_2$ = 0. With increasing $C_2$ = 0.5 $\mu$m, $\lambda$ = 500 nm gives rough agreements for both cases (c and d). However $\lambda$ = 1 $\mu$m does fit neither the first crossing in case A nor the second crossing in case B well. The simulation does not fit the case A anymore with $C_2$ = 10 $\mu$m even though the double crossing in case B can be reproduced (e and f).}
    \label{FigS3}
\end{figure*}

Another potential source of spin currents to consider are interfacial effects at the Pt strips arising from the Kapitza resistance, which creates a temperature discontinuity, $\delta T$ across the YIG$|$Pt interface \cite{cahill2003nanoscale,bender2015interfacial}. One can assume that $\delta T$ is proportional to the temperature gradient at the interface, $\partial_z T$. The measured SSE voltage including the interfacial contribution can be written

\begin{equation}
    V_\text{SSE} = C_1 ( \mu - C_2 k_B\partial_z T),
    \tag{S3}
    \label{eqS3}
\end{equation}
where $\mu$ is the magnon chemical potential. $C_1$ is a constant, which normalizes the simulation results to the experimental data at $d$ = 0 $\mu$m. $C_2$ is a parameter proportional to the Kapitza length, which represents the contribution of the interfacial term. The negative sign implies that the heat flow is along the opposite direction of the temperature gradient. One can assume that $C_2$ is the same for both case A and case B for fixed $d$ because the Al layer does not touch either the YIG or Pt directly. We recall also that in our cartesian frame $z$ is the direction normal to the film.

Figure \ref{FigS3} shows the effect of adding a constant $C_2$ terms for four different values of $\lambda$. For $C_2$ = 0.5 $\mu$m, the simulation can reproduce the observed double crossing in case B for all four values of $\lambda$.  However $\lambda$ = 1 $\mu$m case does not predict well either the observed first crossing in case A or the second crossing in case B. It is important to also point out that $V_\text{SSE}$ eventually follows the temperature gradient profile in Fig. 1(b) of the main text when the $C_2$ term is dominant (Figs. \ref{FigS3}(f)) even for large values of $\lambda$. Thus in the limit of very large $C_2$, the observation of a double crossing of the sign of the SSE in case B is not anymore conspicuous of a short decay length of thermal magnons. But the other consequence of assuming that the interfacial effects are dominant, is to reduce strongly (more than 3 orders of magnitude) the amplitude of the signal after the first crossing in case A. The fact that we observe only a factor 10 reduction of the SSE signal for case A in Fig. 3 of the main text between $d=0$ and $d=0.7$~$\mu$m thus points to a small value of $C_2 \ll 0.5$~$\mu$m.  Experimentally, we have performed an estimation of the Kapitza resistance by comparing the increase of the Pt temperature inferred from the variation  of its resistance and the temperature increase of YIG inferred from the decrease of the Kittel frequency due to a change of $M_s(T)$, whose slope is about 0.4 mT/K at room temperature \cite{thiery2018nonlinear}. We have found no temperature difference between the Pt and the YIG underneath within the uncertainty of 2 K when the increase of temperature rise is $T-T_0=70$~K. At 2 mA, where $T-T_0=130$~K, the temperature gradient is 0.2 K/nm. From this, we estimate an upper bound of Kapitza length of about 20 nm.

\bibliography{bib}